\documentclass[aps,prl,english,superscriptaddress,twocolumn,tightenlines,showpacs,floatfix,amsfonts]{revtex4-1}
\usepackage[T1]{fontenc}
\usepackage[latin9]{inputenc}
\usepackage{xcolor}
\usepackage{textcomp}
\usepackage{mathrsfs}
\usepackage{mathtools}
\usepackage{amsmath}
\usepackage{amssymb}
\usepackage{amsthm}
\usepackage{txfonts}
\usepackage{float}
\usepackage{graphicx}

\usepackage{comment}

\definecolor{heavyred}{rgb}{0.8, 0, 0}

\makeatletter
\@ifundefined{date}{}{\date{}}
\usepackage[colorlinks,bookmarks=false,citecolor=blue,linkcolor=red,urlcolor=blue]{hyperref}

\usepackage{color}
\usepackage{amsmath,amssymb,amsthm,amscd,latexsym,epsfig,bm,subfigure}
\usepackage{mathtools}
\usepackage{leftidx}

\usepackage{babel}

\makeatletter
\let\cat@comma@active\@empty
\makeatother

\makeatother

\usepackage{babel}
\begin{document}
\title{Optimal Thresholds for Fracton Codes and Random Spin Models \\ with Subsystem Symmetry}
\author{Hao Song}
\affiliation{CAS Key Laboratory of Theoretical Physics, Institute of Theoretical Physics, Chinese Academy of Sciences, Beijing 100190, China}
\affiliation{Department of Physics and Astronomy, McMaster University, Hamilton, Ontario L8S 4M1, Canada}
\affiliation{Departamento de F\'isica Te\'orica, Universidad Complutense, 28040 Madrid, Spain}
\author{Janik Sch\"onmeier-Kromer}
\affiliation{Arnold Sommerfeld Center for Theoretical Physics, University of Munich, Theresienstr. 37, 80333 M\"{u}nchen, Germany}
\affiliation{Munich Center for Quantum Science and Technology (MCQST), Schellingstr. 4, 80799 M\"{u}nchen, Germany}
\author{Ke Liu}
\email{ke.liu@lmu.de}
\affiliation{Arnold Sommerfeld Center for Theoretical Physics, University of Munich, Theresienstr. 37, 80333 M\"{u}nchen, Germany}
\affiliation{Munich Center for Quantum Science and Technology (MCQST), Schellingstr. 4, 80799 M\"{u}nchen, Germany}
\author{Oscar Viyuela}
\affiliation{Department of Physics, Massachusetts Institute of Technology, Cambridge, Massachusetts 02139, USA}
\affiliation{Department of Physics, Harvard University, Cambridge, Massachusetts
02318, USA}
\author{Lode Pollet}
\affiliation{Arnold Sommerfeld Center for Theoretical Physics, University of Munich, Theresienstr. 37, 80333 M\"{u}nchen, Germany}
\affiliation{Munich Center for Quantum Science and Technology (MCQST), Schellingstr. 4, 80799 M\"{u}nchen, Germany}
\affiliation{Wilczek Quantum Center, School of Physics and Astronomy, Shanghai Jiao Tong University, Shanghai 200240, China}
\author{M.A. Martin-Delgado}
\email{mardel@fis.ucm.es}
\affiliation{Departamento de F\'isica Te\'orica, Universidad Complutense, 28040
Madrid, Spain}
\date{\today}
\begin{abstract}
Fracton models provide examples of novel gapped quantum phases of matter that host intrinsically immobile excitations and therefore lie beyond the conventional notion of topological order. Here, we calculate optimal error thresholds for quantum error correcting codes based on fracton models. 
By mapping the error-correction process for bit-flip and phase-flip noises into novel statistical models with Ising variables and random multi-body couplings, we obtain models that exhibit an unconventional subsystem symmetry instead of a more usual global symmetry. 
We perform large-scale parallel tempering Monte Carlo simulations to obtain disorder-temperature phase diagrams, which are then used to predict optimal error thresholds for the corresponding fracton code. Remarkably, we found that the X-cube fracton code displays a minimum error threshold ($7.5\%$) that is much higher than 3D topological codes such as the toric code ($3.3\%$), or the color code ($1.9\%$). This result, together with the predicted absence of glass order at the Nishimori line, shows great potential for fracton phases to be used as quantum memory platforms. 
\end{abstract}
\maketitle

\paragraph{Introduction.}
The study of quantum phases constitutes a cornerstone of quantum many-body physics and can potentially enable technological advances. 
Among its modern applications is the possible realization of a fully fledged quantum computer by means of fault-tolerant methods for processing quantum information~\cite{Preskill98,nielsen00,GalindoDelgado}. 
Topological codes~\cite{Kitaev03,ColorCode} stand among the best options to performing fault-tolerant quantum computation due to their high thresholds and linear scaling of the system qubit resources~\cite{Dennis02,CCs2007}.
Nevertheless, 2D topological stabilizer codes like the most studied surface code~\cite{Kitaev03,Dennis02,Bravyi98} permit only topological implementations of Clifford gates~\cite{Bravyi13}, while non-Clifford gates are necessary for realizing the desired quantum advantages~\cite{Aaronson04}, motivating the quest for new 3D codes.
Fracton models~\cite{Chamon05,Bravyi11,Haah11,Yoshida13,Vijay15,
Vijay16,Ma17,VijayNA,Nandkishore19, Song_Twisted,CageNet,
WenDN, WangDN, AasenDN, Devakul18, Muhlhauser20, Zhou22} represent a generalization to 3D  topological orders and provide alternatives to quantum memories beyond the standard paradigm of topological computing.
These models host intrinsically immobile point-like excitations called fractons~\cite{Vijay15} which make a key difference from conventional topological orders and have potential beneficial applications. 
While a few decoders~\cite{RGdecoder,Xcube_EC2019} and several experimental platforms~\cite{Verresen21,Myerson22,You19} are proposed, the theoretical limit on error thresholds of fracton codes is unexplored, which is nevertheless crucial for devising new decoders and for justifying the practical relevance of fracton codes.

The goal of this Letter is to investigate how a fracton model behaves as an active error correcting code against stochastic Pauli errors which are widely used for benchmarking quantum memories. 
We have defined error corrections in the presence of a subextensive ground state degeneracy and computed the optimal thresholds for one of the most representative fracton models in three dimensions---the X-cube model~\cite{Vijay16}. 
We address the problem by a combination of theoretical analyses and numerical simulations.
Using a statistical-mechanical mapping method~\cite{Dennis02} that has previously produced error thresholds for codes beyond those for which it was initially conceived~\cite{Katzgraber09, Bombin12, Katzgraber10, Andrist_2011, Andrist16, Kubica18, Viyuela19, Vodola22},
we derive two statistical models related to Pauli errors of the X-cube model, in the formulation of classical spin variables that are suited for simulations.
The numerical simulation of statistical models in three dimensions with randomness is generally challenging, and the required resources are even higher for our models as they possess  lower-dimensional subsystem symmetries rather than a more conventional global symmetry. 
Only by utilizing state-of-the-art parallel tempering Monte Carlo methods and performing large-scale simulations have we been able to compute various many-body correlation functions and determine the phase diagrams for the two statistical models up to moderate error rates.

We estimate the optimal error thresholds of the X-cube code against  bit-flip ($X$) and phase-flip errors ($Z$) as $p^X_c \simeq 0.152(4)$ and $p^Z_c \simeq 0.075(2)$, respectively. 
The minimum of these thresholds is remarkably higher than what was found in conventional 3D topological codes such as the toric code ($0.033$)~\cite{Wang02, Ohno2004} and the color code ($0.019$)~\cite{Kubica18}, 
which signals the potential of the X-cube model as a fault-tolerant quantum memory. This is further confirmed by the analytical result that the Nishimori line is free of fracton glass order through which the resilience of the quantum code may be lost (see SM~\cite{SM}). 
In addition, our results represent the first study of spin models with both subsystem symmetries and quenched random disorder in three dimensions, hence are also of interest for the statistical mechanics community.

\begin{figure}[t]
\includegraphics[width=1\columnwidth]{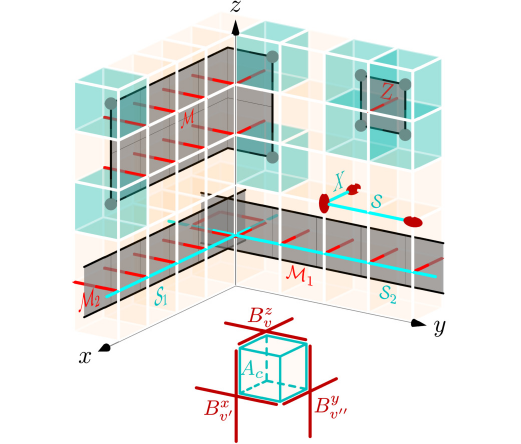}
\caption{Stabilizer generators, excitations, and logical operators of the X-cube model. 
A physical qubit is assigned to each edge ($\ell$) of the lattice.
A Pauli $X$ ($Z$) operator is presented as a cyan (red) edge.
Stabilizer generators $A_{c}$ and $B_{v}^{\mu}$ are defined at unit cubes ($c$) and vertices ($v$), respectively.
The ground states satisfy $A_{c} = 1$ and $B^{\mu}_v = 1$ for all $c, v$ and $\mu \in \{x, y, z\}$.
Cyan strings ($\mathcal{S}$) and gray membranes ($\mathcal{M}$) represent string and membrane operators of the form $X_\mathcal{S} \coloneqq \prod_{\ell \in \mathcal{S}} X_\ell$ 
and $Z_\mathcal{M} \coloneqq \prod_{\ell \cap \mathcal{M} \neq \emptyset} Z_\ell$, respectively.
Fracton excitations $A_{c} = -1$ are created by membrane operators at their corners and colored as cyan cubes. String operators create lineon excitations at their turns and ends.
The red ellipsoids represent lineons with $B^{\mu}_v = 1, \, B^{\nu\neq \mu}_v = -1$, where $\mu$ is indicated by the elongated directions.
The extended strings and membranes  $\mathcal{S}_1$, $\mathcal{S}_2$, $\mathcal{M}_1$, and $\mathcal{M}_2$ represent two pairs of logical operators creating no excitations due to PBC.}
\label{fig:Xcube_errors}
\end{figure}

\paragraph{X-cube model as quantum memory.}
Consider a cubic lattice $\mathcal{L}$ with periodic boundary condition (PBC).
We introduce a qubit to every edge $\ell$ and define stabilizer generators $A_c$ and $B^{\mu}_v$ at each unit cube $c$ and vertex $v$ of the lattice.
Specifically, $A_c$ is defined to be the tensor product of Pauli $X$ operators on the $12$ edges of a cube, and $B^{\mu}_v$ is a tensor product of Pauli $Z$ operators on the four edges adjacent to a vertex and perpendicular to the spacial direction $\mu = x,y,z$.
Namely,
\begin{align}\label{eq:stabilizers}
	A_{c}\coloneqq \prod_{\ell\in c} X_{\ell},
	\qquad B_{v}^{\mu}\coloneqq \prod_{\ell\in v:\ell\perp\mu}Z_{\ell},
\end{align}
as visualized in Fig.~\ref{fig:Xcube_errors}.
For convenience, we label the set of qubits as $\mathcal{Q}$ and those of stabilizer generators as $\mathcal{A} \coloneqq \{A_c\}$ and $\mathcal{B} \coloneqq \{B^x_v, B^y_v\}$, respectively.

The X-cube model is a paradigmatic fracton model constructed by summing over these stabilizer generators,
\begin{align}\label{eq:X-cube}
H_{\text{X-cube}}=-\sum_{c \in \mathcal{L}}A_{c}-\sum_{v \in \mathcal{L}}\left(B_{v}^{x}+B_{v}^{y}+B_{v}^{z}\right).
\end{align}
As $A_c$ and $B^{\mu}_v$ commute, this Hamiltonian is exactly solvable~\cite{Vijay16}, and its ground states satisfy $A_c = 1$ and $B^{\mu}_v = 1$ for all $c, \, v, \, \mu$.
The elementary excitations are two types of gapped topological defects.
A unit cube with $A_c = -1$ is referred to as a \emph{fracton} (solid cyan cube in Fig.~\ref{fig:Xcube_errors}), which is an intrinsically \emph{immobile} defect.
A vertex with $B^{\mu}_v = 1$ but $B^{\nu\neq \mu}_v = -1$ corresponds to an excitation termed a \emph{lineon} (red ellipsoid in Fig.~\ref{fig:Xcube_errors}), which can move along the $\mu$ direction but is immobile in the two directions $\nu \neq \mu$.
The three possible lineons at each vertex are subject to a constraint $B^x_v B^y_v B^z_v = 1$.

On a lattice of size $L^3$ with PBC, the X-cube model has $2^{6L-3}$ degenerated ground states, scaling subextensively with system size~\cite{Vijay16,Song_Twisted}. 
These ground states are indistinguishable by local operations, hence provide a fault-tolerant code Hilbert space.
We can view them as $6L-3$ logical quibits by introducing $6L-3$ pairs of non-local operators $(X_{\mathcal{S}_j}, Z_{\mathcal{M}_j})$.
Here, $X_{\mathcal{S}_j} \coloneqq \prod_{\ell \in \mathcal{S}_j} X_\ell$ and $Z_{\mathcal{M}_j} \coloneqq \prod_{\ell \cap \mathcal{M}_j \neq \emptyset} Z_\ell$ are defined on \emph{extended} strings and membranes winding around the lattice (see Fig.~\ref{fig:Xcube_errors}).

\paragraph{Error correction.}
Fractons and lineons can be used to diagnose errors in the stabilizer code.
For example, as illustrated in Fig.~\ref{fig:Xcube_errors}, a phase-flip $Z$ error on a single qubit $\ell$ will cause four fractons at each of its adjacent cubes. Similarly, a single bit-flip $X$ error will create two lineons at the vertices sharing the edge.
Therefore, an ensemble of fractons or lineons can act as an $A$- or $B$-syndrome reflecting $Z$ or $X$ errors in the system.

For simplicity, we consider a situation where each qubit is affected by phase-flip and bit-flip errors independently and assume perfect measurements for all stabilizer generators.
Moreover, since the X-cube model is a Calderbank-Shor-Steane (CSS) code~\cite{CSS}, i.e., the type-$A$ and type-$B$ stabilizer generators involve either purely $X$ or $Z$ operators, we can correct the bit-flip and phase errors separately.

The error-correction process can be described by introducing a (co)chain complex,
\begin{align}\label{eq:chain}
\begin{array}{ccccc}
\mathbb{Z}_{2}^{\mathcal{A}} & \xrightleftharpoons[\partial_{A}^{\dagger}]{\partial_{A}} & \mathbb{Z}_{2}^{\mathcal{Q}} & \xrightleftharpoons[\partial_{B}]{\partial_{B}^{\dagger}} & \mathbb{Z}_{2}^{\mathcal{B}}\end{array},
\end{align}
where $\mathbb{Z}_{2}^{\mathcal{A}}$, $\mathbb{Z}_{2}^{\mathcal{Q}}$ and $\mathbb{Z}_{2}^{\mathcal{B}}$ denote the $\mathbb{Z}_{2} = \{0, 1\}$ vector spaces for labeling configurations of type-$A$ stabilizer generators ($\mathcal{A}$), physical qubits ($\mathcal{Q}$), and type-$B$ stabilizer generators ($\mathcal{B}$), respectively.
The boundary maps $\partial_{A}$ and $\partial_{B}$ are linear and specify the qubits involved in every \mbox{type-$A$} and type-$B$ stabilizer generator.
Correspondingly, the transpose operator $\partial_{A}^{\dagger}$ ($\partial_{B}^{\dagger}$) maps an error configuration $\eta \in \mathbb{Z}_{2}^{\mathcal{A}(\mathcal{B})}$ to an ensemble of fractons (lineons) 
created by $Z_{\eta} \coloneqq \prod_{\ell \in \eta} Z_{\ell}$ 
($X_{\eta} \coloneqq \prod_{\ell \in \eta} X_{\ell}$).
In general, $\partial_{A}^{\dagger}\circ\partial_{B}=\partial_{B}^{\dagger}\circ\partial_{A}=0$ for all CSS codes.

Only certain error configurations are compatible with a given syndrome. Among those, error configurations are equivalent if and only if they can be connected by \mbox{type-$A$} and type-$B$ stabilizer generators.
Namely, provided $\eta - \eta^\prime \in \mathrm{im\,}\partial_{A}$ or $\mathrm{im\,}\partial_{B}$, two errors $\eta$ and $ \eta^\prime$ will have the same effect on the encoded quantum state, where $\mathrm{im\,}\partial$ denotes the image of the boundary map.
Thus, the spaces of $X_\eta$ and $Z_\eta$ can be divided into equivalence classes by the quotients 
$\mathbb{Z}_{2}^{\mathcal{Q}}/\mathrm{im\,}\partial_{A}$ 
and  $\mathbb{Z}_{2}^{\mathcal{Q}}/\mathrm{im\,}\partial_{B}$. 
We denote the equivalence classes as 
$\left[\eta\right]_{X}\coloneqq\eta+\mathrm{im\,}\partial_{A}$ 
and $\left[\eta\right]_{Z}\coloneqq\eta+\mathrm{im\,}\partial_{B}$, respectively.

For a possible $B$- or $A$-syndrome $\sigma$ with probability $\mathrm{Pr}(\sigma)$, the total probability of those equivalence classes compatible with $\sigma$ satisfies $\sum_{\lambda}\mathrm{pr}\left(\left[\eta_{\sigma} + \lambda \right]\right) = \mathrm{Pr}(\sigma)$,
with $\lambda \in \mathbb{Z}_{2}^{\mathcal{Q}}$ labeling inequivalent logical operators.
The correction can be realized successfully if, for typical syndromes, there exists a \emph{most probable} equivalence class $\left[\eta^{*}_{\sigma}\right]$ such that $\mathrm{pr}\left(\left[\eta^*_{\sigma}\right]\right) \rightarrow \mathrm{Pr}(\sigma)$ in the large system limit~\cite{Dennis02}.
However, this is only possible when the rate $p$ for local $X$ and $Z$ errors lie below some optimal threshold values $p^X_c$ and $p^Z_c$.
For $p > p_c$, the error class cannot be unambiguously identified, and the code becomes ineffective.
Finding the optimal error thresholds is therefore crucial to any quantum code.

\begin{figure}
\centering
\includegraphics[width=0.45\textwidth]{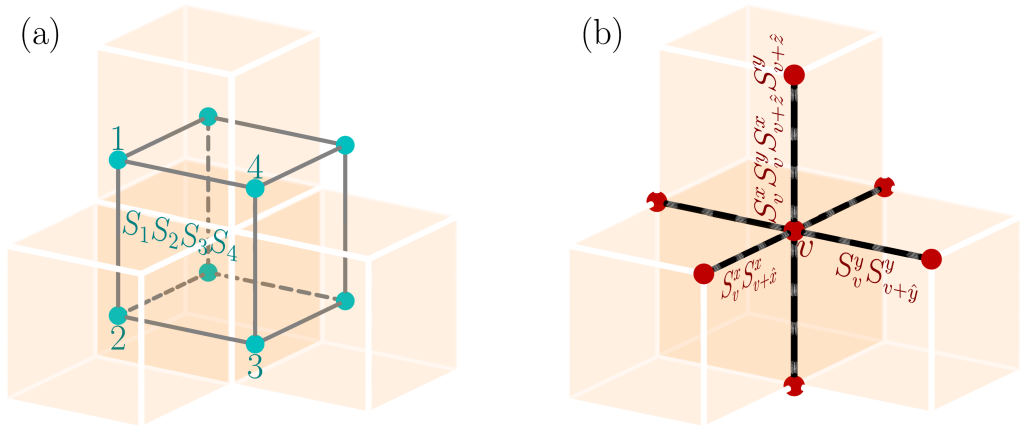}
\caption{Illustration of the 3D random plaquette Ising (RPI) model and the 3D random anisotropically coupled Ashkin-Teller (RACAT) model.
({\bf a}) The four interacting $S_c$ spins in the RPI model correspond to the four type-$A$ stabilizer generators sharing an edge in the original X-cube model.
({\bf b}) 
	The RACAT model has two spins $S_{v}^{x}$ and $S_{v}^{y}$ at each vertex, corresponding the two independent type-$B$ stabilizer generators.
The coupling coefficients take random signs depending on the absence and presence of local errors.}
\label{fig:Xcube_statM}
\end{figure}

\begin{figure*}[t]
\centering
\includegraphics[width=1\textwidth]{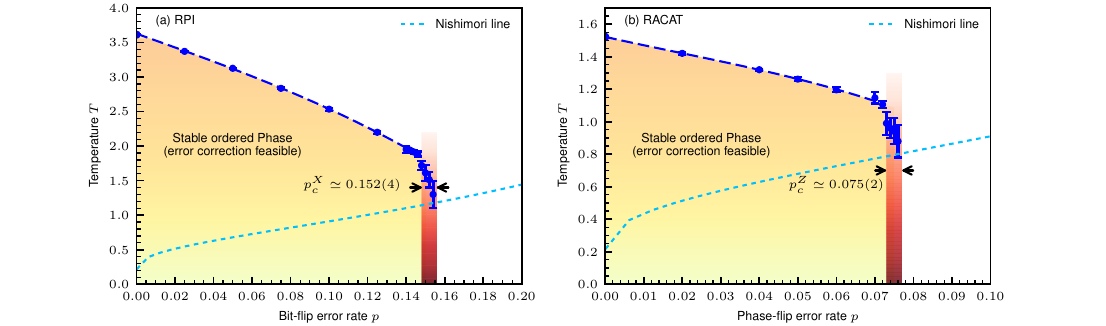}
\caption{
Phase diagrams of the 3D RPI model and 3D RACAT model.
The phase transitions are discontinuous in the low error rate ($p$) regions (dashed lines) but are softened to be continuous ones when approaching the threshold values $p^X_c \simeq 0.152(4)$ and $p^Z_c \simeq 0.075(2)$.
The error thresholds are determined by the largest $p$ values exhibiting an order-disorder phase transition (see SM~\cite{SM}).
The effective temperature $T$ is an auxiliary quantity and related to $p$ by the Nishimori line (dotted line).
A correctable X-cube code corresponds to the part of the Nishimori line inside the ordered phases.}
\label{fig:PD}
\end{figure*}

\paragraph{Mapping to statistical-mechanical models.}
An elegant and numerically preferable way to determine $p^X_c$ and $p^Z_c$ of the X-cube code is utilizing a statistical mapping method~\cite{Dennis02} which maps bit- and phase-flip errors to suitably chosen statistical-mechanical models.

Suppose both $X$ and $Z$ errors occur independently at each qubit at rate $p$.
Then the probability the system is affected by an $X$ or $Z$ error configuration $\eta \in \mathbb{Z}_2^{\mathcal{Q}}$ is
\begin{equation}\label{eq:pr}
\mathrm{pr}\left(\eta;p\right)=\prod_{\ell\in\mathcal{Q}}p^{\eta\left(\ell\right)}\left(1-p\right)^{1- \eta\left(\ell\right)}\propto\left(\frac{p}{1-p}\right)^{\sum_{\ell}\eta\left(\ell\right)},
\end{equation}
where $\eta(\ell) = 1$ or $0$ on edges with or without an error.

This probability can be interpreted as a Boltzmann weight by introducing an effective temperature $T$ satisfying 
\begin{equation}\label{eq:Nishimori}
e^{-\frac{2}{T}}=\frac{p}{1-p}.
\end{equation}
Eq.~\eqref{eq:Nishimori} defines the so-called \emph{Nishimori line} and allows us to control the rate of random qubit errors through the auxiliary temperature (see SM~\cite{SM}).

Accordingly, the total probability of a bit-flip error equivalence class $\left[\eta\right]_{X} \coloneqq \eta+\mathrm{im\,}\partial_{A}$
is mapped to the partition function of an interacting spin Hamiltonian $H^{\mathcal{A}}_{\eta}$,
\begin{align}\label{eq:Xpartition}
\text{pr}\left(\left[\eta\right]_{X};p\right) & \propto \sum_{f\in\mathbb{Z}_{2}^{\mathcal{A}}} e^{\beta \sum_{\ell \in \mathcal{Q}}\left(-1\right)^{\eta (\ell) + \partial_{A} f (\ell)}} \nonumber \\
& = \sum_{\left\{ S_{c}=\pm1\right\} }e^{-\beta H_{\eta}^{\mathcal{A}}\left(\left\{ S_{c}\right\} \right)} \eqqcolon \mathcal{Z}_{\eta}^{\mathcal{A}}\left(\beta\right),
\end{align}
where $\beta=1/T$ is the inverse temperature, 
$f \equiv \{f(c)\}_{c\in\mathcal{A}} \in \mathbb{Z}_2^{\mathcal{A}}$ represents a configuration of type-$A$ stabilizer generators,
$\partial_{A}f (\ell) = \sum_{c\in \partial_{A}^{\dagger}\ell} f(c)$ labels the edges of $f$, and $S_c = (-1)^{f(c)} \in \{\pm 1 \}$ denotes effective Ising variables on the center of cubes.

The form of $H^{\mathcal{A}}_{\eta}$ realizes a 3D \emph{random plaquette Ising} (RPI) model on a dual lattice with quenched disorder,
\begin{align}\label{eq:Hx}
H_{\eta}^{\mathcal{A}}\left(\left\{ S_{c}\right\} \right)=-\sum_{\ell \in \mathcal{Q}}\left(-1\right)^{\eta\left(\ell\right)}\prod_{c\in\partial_{A}^{\dagger}\ell}S_{c},
\end{align} 
where $c \in \partial_{A}^{\dagger}\ell$ specifies the four type-$A$ stabilizer generators sharing an edge $\ell$ (see Fig.~\ref{fig:Xcube_statM}),
while the probabilities of their coupling to be anti-ferromagnetic and ferromagnetic are $p$ and $1-p$, respectively.
Moreover, aside from a global $Z_2$ symmetry, $H^{\mathcal{A}}_{\eta}$ is invariant under a subsystem symmetry flipping spins in individual planes and may be viewed as a novel random spin model.

Analogously, the modeling of a phase-flip error equivalence class $\left[\eta\right]_{Z}\coloneqq\eta+\mathrm{im\,}\partial_{B}$ leads to a 3D \emph{random anisotropically coupled Ashkin-Teller} (RACAT) model with quenched disorder on the original cubic lattice, 
\begin{align}\label{eq:Hz}
H_{\eta}^{\mathcal{B}}\left(\left\{ S_{v}^{\mu}\right\}\right)=-\sum_{v}\sum_{\mu=x,y,z}\left(-1\right)^{\eta\left(\ell_{v}^{\hat{\mu}}\right)}S_{v}^{\mu}S_{v+\hat{\mu}}^{\mu},
\end{align}
where $S_{v}^{x}=\left(-1\right)^{f_{v}^{y}}$, $S_{v}^{y}=\left(-1\right)^{f_{v}^{x}}$, and $S_{v}^{z} = S_{v}^{x} S_{v}^{y}$
are effective Ising variables, and
$f\equiv\left\{ f_{v}^{x},f_{v}^{y}\right\} _{v}\in\mathbb{Z}_{2}^{\mathcal{B}}$
denotes the indicator vectors for type-$B$ stabilizer generators.
The spins are coupled only along with the unit $\hat{\mu}$ direction (Fig.~\ref{fig:Xcube_statM}).
In contrast to the usual 3D Ashkin-Teller model~\cite{Kadanoff80}, the RACAT model Eq.~\eqref{eq:Hz} has the planar symmetries of flipping all $S^{\mu}$ and $S^{\nu}$ spins in an arbitrary $\mu$-$\nu$ plane besides a global $Z_2 \times Z_2$ symmetry.

The disorder-free limits ($p=0$) of $H_{\eta}^{\mathcal{A}}$ and $H_{\eta}^{\mathcal{B}}$ are dual to each other~\cite{Johnston11}, as for general fracton and topological CSS codes~\cite{SM}. 
There is no exact duality in the presence of disorder, nevertheless, our results  suggest an approximate duality relation between  the error thresholds $p^X_c$ and $p^Z_c$~\cite{SM}.
 
On the side of the statistical-mechanical models, the relative probability between two $X$ (or $Z$) error equivalence classes under the error rate $p$ is given by the difference between their free energies, 
\begin{align}\label{eq:deltaF}
\frac{\text{pr}\left(\left[\eta+\lambda\right]_{X(Z)};p\right)}{\text{pr}\left(\left[\eta\right]_{X(Z)};p\right)}=\frac{\mathcal{Z}_{\eta+\lambda}^{\mathcal{A}(\mathcal{B})}\left(\beta\right)}{\mathcal{Z}_{\eta}^{\mathcal{A}(\mathcal{B})}\left(\beta\right)}=e^{-\beta\delta\mathcal{F}_{\eta,\lambda}^{\mathcal{A}(\mathcal{B})}},
\end{align}
where $\lambda \in\mathbb{Z}_{2}^{\mathcal{Q}}$ represents logical $X$ (or $Z$) operators of the X-cube model and can flip a sequence of coupling coefficients in $H_{\eta}^{\mathcal{A}}$ ($H_{\eta}^{\mathcal{B}}$).
The condition of existing 
$\mathrm{pr}\left(\left[\eta^{*}_{\sigma}\right]_{X(Z)}\right) \rightarrow \mathrm{Pr}(\sigma)$
for the \emph{most probable} equivalence class $\left[\eta^{*}_{\sigma}\right]_{X(Z)}$ requires that the free energy to introduce a nontrivial string (membrane) defect ($\lambda \neq 0$) diverges in the thermodynamical limit, namely,
$\delta\mathcal{F}_{\eta,\lambda}^{\mathcal{A}(\mathcal{B})}\rightarrow\infty$ (see SM~\cite{SM}).
This is only possible when $H_{\eta}^{\mathcal{A}}$ and $H_{\eta}^{\mathcal{B}}$ are in their ordered phases.
Hence $p^X_c$ and $p^Z_c$ can be determined from the order-disorder phase transitions of the two random spin models.

\begin{figure}[t]
\centering
\includegraphics[width=0.5\textwidth]{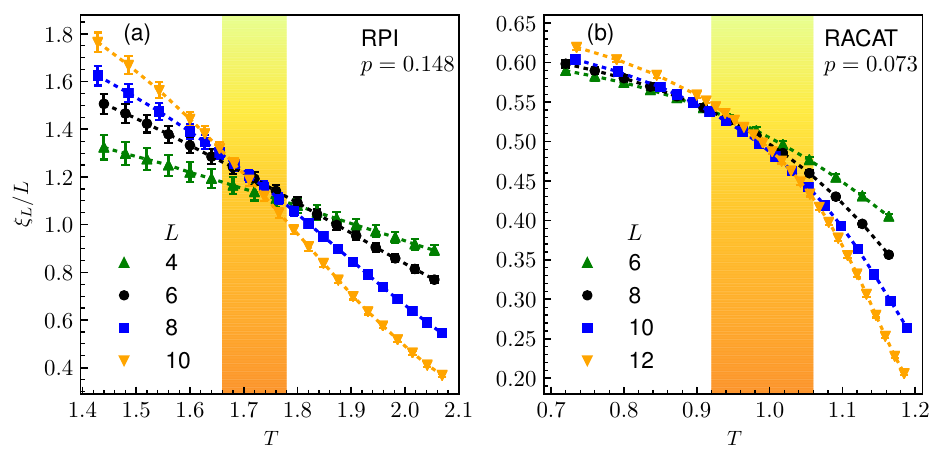}
 \caption{Normalized second-moment correlation length $\xi_L/L$ for the 3D RPI (a) and RACAT (b) model close to the error thresholds $p_c^X \simeq 0.152(4)$ and $p_c^Z \simeq 0.075(2)$.
 For a second-order phase transition the curves for different system sizes intersect near the critical temperature $T_c$. The shaded areas show a conservative estimate of $T_c$.
Such intersections cannot be recognized for $p > p_c^X$ or $p_c^Z$ where we conclude no phase transition.}
 \label{fig:xi}
 \end{figure}

\paragraph{Error thresholds and phase diagrams.}
The phase diagrams of the RPI and RACAT models are shown in Fig.~\ref{fig:PD}, obtained by large-scale parallel tempering Monte Carlo simulations.
We locate the phase transitions by cross-checking the energy histogram, specific heat, order parameter,  its susceptibility, and the correlation length~\cite{SM}.

To construct the appropriate order parameters, the planar symmetries of $H_{\eta}^{\mathcal{A}}$ and $H_{\eta}^{\mathcal{B}}$ have to be taken into account as they can lead to trivial cancellation of local orders.
For the RPI model, we define 
\begin{align}\label{eq:opA}
	Q^{\mathcal{A}} \coloneqq \frac{1}{L^3} \sum^{L-1}_{z=0} \left[\left\langle\left|\sum^{L-1}_{x,y=0} S_{c(x,y,z)}S_{c(x,y,z+1)}	\right| \right\rangle \right],
\end{align} 
with $\langle . \rangle$ and $[.]$ denoting the thermal and disorder average, respectively.
The inner sum in Eq.~\eqref{eq:opA} involves a subextensive number ($\propto L^2$) of spins, while the norm enforces the planar-flip invariance~\cite{Johnston17}.
Thus, $Q^{\mathcal{A}}$ defines a long-range order which is \emph{sub-dimensional} and made of \emph{plane-like} objects.

The order parameter for the RACAT model is constructed similarly
\begin{align} \label{eq:opB}
	Q^{\mathcal{B}} \coloneqq \frac{1}{L^3} \sum^{L-1}_{x,y = 0}\left[\left\langle \left|\sum^{L-1}_{z = 0} S^{z}_{v(x,y,z)}	\right| \right\rangle \right],
\end{align}
which describes an order for extended \emph{line} objects.
Here, the $S^z_v$ spins are taken for simplicity, as the three Ising variables in Eq.~\eqref{eq:Hz} are permutable.

At low $X$ and $Z$ error rates, the energy histograms reveal a first-order phase transition for both $H_{\eta}^{\mathcal{A}}$ and $H_{\eta}^{\mathcal{B}}$~\cite{SM}.
This agrees with previous studies on the disorder-free ($p=0$) limit of the two models~\cite{Johnston11, Johnston17}. Hence the transition temperatures can be estimated in a relatively straightforward way~\cite{SM}.

For larger $p$, the phase transitions are softened to continuous ones in line with the Imry-Ma scenario~\cite{Imry75, Imry79}.
We can then locate the transitions by studying the second-moment correlation length
\begin{align}\label{eq:xi}
	\xi_L \coloneqq \frac{1}{2\sin\left(\lvert\mathbf{k}_{\min}\rvert/2\right)}\left(\frac{\tilde{G}\left(\mathbf{0}\right)}{\tilde{G}\left(\mathbf{k}_{\min}\right)}-1\right)^{1/2},
\end{align}
where $\tilde{G}\left(\mathbf{k}\right)\coloneqq\sum_{\mathbf{r}}G\left(\mathbf{r}\right)e^{-i\mathbf{k}\cdot\mathbf{r}}$
is a Fourier transform of the spatial correlator $G\left(\mathbf{r}\right)$, and $\mathbf{k}_{\min}$ denotes any smallest non-zero wave vector~\cite{Janke08}.

The spatial correlators related to the order parameters $Q^{\mathcal{A}}$ and $Q^{\mathcal{B}}$ are given by
\begin{align}
	&G^{\mathcal{A}}\left(\mathbf{r}\right) \coloneqq\frac{1}{L^{3}}\sum_{c \in \mathcal{L}}\left[\left\langle S_{c}S_{c+\hat{z}}S_{c+\left(\mathbf{r},0\right)}S_{c+\left(\mathbf{r},\hat{z}\right)}\right\rangle\right], \label{eq:GA} \\
	&G^{\mathcal{B}}\left(r\right) \coloneqq\frac{1}{L^3}\sum_{v \in \mathcal{L}}\left[\left\langle S_{v}^{z}S_{v+r\hat{z}}^{z}\right\rangle\right],\label{eq:GB} 
	\end{align}
so that $G^{\mathcal{A}(\mathcal{B})}\left(\mathbf{r}\right) \rightarrow \left[Q^{\mathcal{A}(\mathcal{B})}\right]^2$
in the limit of $|\mathbf{r}| \rightarrow \infty$.

Provided a continuous phase transition exists, $\xi_L/L$ is expected to be scaled as $g\left(L^{1/\nu}\left(T-T_{c}\right)\right)$ near the critical point, and the curves for different system sizes should intersect at $\left(T_{c},g\left(0\right)\right)$,
where $g$ is a universal scaling function and $\nu$ denotes the critical exponent of $\xi$. 
We can use this property to search the largest error rates where $H_{\eta}^{\mathcal{A}}$ and $H_{\eta}^{\mathcal{B}}$ continue showing a continuous phase transition, which in turn implies the error thresholds of the X-cube code.

Our simulations show that the $\xi_L(T)/L$ curves exhibit clear intersections up to an error rate $p^X_c \simeq 0.152(4)$ for the RPI model and $p^Z_c \simeq 0.075(2)$ for the RACAT model (see Fig.~\ref{fig:xi} and also SM~\cite{SM}).
Thereafter, a clear intersection cannot be recognized~\cite{SM}, indicating lack of an order-disorder transition.
Namely, for error rates larger than $p^X_c$ and $p^Z_c$, $H_{\eta}^{\mathcal{A}}$ and $H_{\eta}^{\mathcal{B}}$ host no long-range order, and the X-cube code hence becomes uncorrectable.
 
\paragraph{Conclusions.}
The X-cube model is the archetypal stabilizer code exhibiting the fascinating quantum physics of fracton topological orders.
In this work we investigated its capability as a quantum memory through a combination of theoretical analyses and detailed numerical simulations.
We estimated its optimal error thresholds as  $p^X_c \simeq 15.2(4)\%$ against bit-flip noise and $p^Z_c \simeq 7.5(2)\%$ against phase-flip noise, featuring a remarkably higher minimum error rate ($7.5\%$) compared to the 3D toric code ($3.3\%$)~\cite{Wang02, Ohno2004} and color code ($1.9\%$)~\cite{Kubica18}.
Our work establishes the general connection between the fault tolerance of fracton codes and statistical-mechanical models with subsystem symmetries. 
The Pauli error thresholds in any CSS code with zero-encoding rate~\cite{Dennis02, Katzgraber09, Bombin12, Kubica18} obey the inequality $H\left(p_{c}^{X}\right)+H\left(p_{c}^{Z}\right)\leq1$ imposed by the quantum Gilbert-Varshamov bound~\cite{Gilbert52, Varshamov57, Calderbank96}, 
where  $H\left(p\right)\coloneqq-p\log_{2}\left(p\right)-\left(1-p\right)\log_{2}\left(1-p\right)$ is the Shannon entropy.
Our results give  $H\left(p_{c}^{X}\right)+H\left(p_{c}^{Z}\right) \simeq 1.00(2)$,  which not only satisfies this constraint but is close to its upper bound, similar to the situations found in conventional topological codes~\cite{Katzgraber09,Kubica18,Nishimori2007}.
We formulate this near saturation via an approximate duality in the SM~\cite{SM} and conjecture it for general fracton and topological CSS codes.
Our work can guide further studies of fracton models, and the approximate duality predicts even higher thresholds for the self-dual checkerboard~\cite{Vijay16} and Haah's~\cite{Haah11} codes.

\begin{acknowledgments}
H.S. and M.A.M.-D. acknowledge support from the Spanish MINECO grants MINECO/FEDER Projects FIS2017-91460-EXP, PGC2018-099169-B-I00 FIS-2018, and with O.V. from CAM/FEDER Project No.S2018/TCS-4342 (QUITEMAD-CM).
H.S. has also been supported by the Natural Sciences and Engineering Research Council of Canada and the National Natural Science Foundation of China (Grant No.~12047503).
M.A.M.-D. has also been supported by MCIN with funding from European Union NextGenerationEU (PRTR-C17.I1) and the Ministry of Economic Affairs Quantum ENIA project, and partially by the U.S.Army Research Office through Grant No. W911NF-14-1-0103.
J. S.-K., K.L., and L.P. acknowledge support from FP7/ERC Consolidator Grant QSIMCORR, No. 771891, and the Deutsche Forschungsgemeinschaft (DFG, German Research Foundation) under Germany's Excellence Strategy -- EXC-2111 -- 390814868.
The project/research is part of the Munich Quantum Valley, which is supported by the Bavarian state government with funds from the Hightech Agenda Bayern Plus.
Our simulations make use of the ALPS\-Core library~\cite{Gaenko17} and the TK\-SVM library~\cite{Greitemann19, Liu19}.
The data used in this work are available in Ref.~\cite{data}.
\end{acknowledgments}

\bibliographystyle{apsrev4-1}
\bibliography{XcubeEC}

\begin{thebibliography}{68}%
\makeatletter
\providecommand \@ifxundefined [1]{%
 \@ifx{#1\undefined}
}%
\providecommand \@ifnum [1]{%
 \ifnum #1\expandafter \@firstoftwo
 \else \expandafter \@secondoftwo
 \fi
}%
\providecommand \@ifx [1]{%
 \ifx #1\expandafter \@firstoftwo
 \else \expandafter \@secondoftwo
 \fi
}%
\providecommand \natexlab [1]{#1}%
\providecommand \enquote  [1]{``#1''}%
\providecommand \bibnamefont  [1]{#1}%
\providecommand \bibfnamefont [1]{#1}%
\providecommand \citenamefont [1]{#1}%
\providecommand \href@noop [0]{\@secondoftwo}%
\providecommand \href [0]{\begingroup \@sanitize@url \@href}%
\providecommand \@href[1]{\@@startlink{#1}\@@href}%
\providecommand \@@href[1]{\endgroup#1\@@endlink}%
\providecommand \@sanitize@url [0]{\catcode `\\12\catcode `\$12\catcode
  `\&12\catcode `\#12\catcode `\^12\catcode `\_12\catcode `\%12\relax}%
\providecommand \@@startlink[1]{}%
\providecommand \@@endlink[0]{}%
\providecommand \url  [0]{\begingroup\@sanitize@url \@url }%
\providecommand \@url [1]{\endgroup\@href {#1}{\urlprefix }}%
\providecommand \urlprefix  [0]{URL }%
\providecommand \Eprint [0]{\href }%
\providecommand \doibase [0]{http://dx.doi.org/}%
\providecommand \selectlanguage [0]{\@gobble}%
\providecommand \bibinfo  [0]{\@secondoftwo}%
\providecommand \bibfield  [0]{\@secondoftwo}%
\providecommand \translation [1]{[#1]}%
\providecommand \BibitemOpen [0]{}%
\providecommand \bibitemStop [0]{}%
\providecommand \bibitemNoStop [0]{.\EOS\space}%
\providecommand \EOS [0]{\spacefactor3000\relax}%
\providecommand \BibitemShut  [1]{\csname bibitem#1\endcsname}%
\let\auto@bib@innerbib\@empty
\bibitem [{\citenamefont {Preskill}(1998)}]{Preskill98}%
  \BibitemOpen
  \bibfield  {author} {\bibinfo {author} {\bibfnamefont {J.}~\bibnamefont
  {Preskill}},\ }\enquote {\bibinfo {title} {Fault-tolerant quantum
  computation},}\ in\ \href {\doibase 10.1142/9789812385253_0008} {\emph
  {\bibinfo {booktitle} {Introduction to Quantum Computation and
  Information}}}\ (\bibinfo  {publisher} {World Scientific},\ \bibinfo {year}
  {1998})\ pp.\ \bibinfo {pages} {213--269}\BibitemShut {NoStop}%
\bibitem [{\citenamefont {Nielsen}\ and\ \citenamefont
  {Chuang}(2000)}]{nielsen00}%
  \BibitemOpen
  \bibfield  {author} {\bibinfo {author} {\bibfnamefont {M.~A.}\ \bibnamefont
  {Nielsen}}\ and\ \bibinfo {author} {\bibfnamefont {I.~L.}\ \bibnamefont
  {Chuang}},\ }\href@noop {} {\emph {\bibinfo {title} {Quantum Computation and
  Quantum Information}}}\ (\bibinfo  {publisher} {Cambridge University Press},\
  \bibinfo {year} {2000})\BibitemShut {NoStop}%
\bibitem [{\citenamefont {Galindo}\ and\ \citenamefont
  {Mart\'{\i}n-Delgado}(2002)}]{GalindoDelgado}%
  \BibitemOpen
  \bibfield  {author} {\bibinfo {author} {\bibfnamefont {A.}~\bibnamefont
  {Galindo}}\ and\ \bibinfo {author} {\bibfnamefont {M.~A.}\ \bibnamefont
  {Mart\'{\i}n-Delgado}},\ }\href {\doibase 10.1103/RevModPhys.74.347}
  {\bibfield  {journal} {\bibinfo  {journal} {Rev. Mod. Phys.}\ }\textbf
  {\bibinfo {volume} {74}},\ \bibinfo {pages} {347} (\bibinfo {year}
  {2002})}\BibitemShut {NoStop}%
\bibitem [{\citenamefont {Kitaev}(2003)}]{Kitaev03}%
  \BibitemOpen
  \bibfield  {author} {\bibinfo {author} {\bibfnamefont {A.}~\bibnamefont
  {Kitaev}},\ }\href {\doibase https://doi.org/10.1016/S0003-4916(02)00018-0}
  {\bibfield  {journal} {\bibinfo  {journal} {Annals of Physics}\ }\textbf
  {\bibinfo {volume} {303}},\ \bibinfo {pages} {2 } (\bibinfo {year}
  {2003})}\BibitemShut {NoStop}%
\bibitem [{\citenamefont {Bombin}\ and\ \citenamefont
  {Martin-Delgado}(2006)}]{ColorCode}%
  \BibitemOpen
  \bibfield  {author} {\bibinfo {author} {\bibfnamefont {H.}~\bibnamefont
  {Bombin}}\ and\ \bibinfo {author} {\bibfnamefont {M.~A.}\ \bibnamefont
  {Martin-Delgado}},\ }\href {\doibase 10.1103/PhysRevLett.97.180501}
  {\bibfield  {journal} {\bibinfo  {journal} {Phys. Rev. Lett.}\ }\textbf
  {\bibinfo {volume} {97}},\ \bibinfo {pages} {180501} (\bibinfo {year}
  {2006})}\BibitemShut {NoStop}%
\bibitem [{\citenamefont {Dennis}\ \emph {et~al.}(2002)\citenamefont {Dennis},
  \citenamefont {Kitaev}, \citenamefont {Landahl},\ and\ \citenamefont
  {Preskill}}]{Dennis02}%
  \BibitemOpen
  \bibfield  {author} {\bibinfo {author} {\bibfnamefont {E.}~\bibnamefont
  {Dennis}}, \bibinfo {author} {\bibfnamefont {A.}~\bibnamefont {Kitaev}},
  \bibinfo {author} {\bibfnamefont {A.}~\bibnamefont {Landahl}}, \ and\
  \bibinfo {author} {\bibfnamefont {J.}~\bibnamefont {Preskill}},\ }\href
  {\doibase 10.1063/1.1499754} {\bibfield  {journal} {\bibinfo  {journal}
  {Journal of Mathematical Physics}\ }\textbf {\bibinfo {volume} {43}},\
  \bibinfo {pages} {4452} (\bibinfo {year} {2002})}\BibitemShut {NoStop}%
\bibitem [{\citenamefont {Bombin}\ and\ \citenamefont
  {Martin-Delgado}(2007)}]{CCs2007}%
  \BibitemOpen
  \bibfield  {author} {\bibinfo {author} {\bibfnamefont {H.}~\bibnamefont
  {Bombin}}\ and\ \bibinfo {author} {\bibfnamefont {M.~A.}\ \bibnamefont
  {Martin-Delgado}},\ }\href {\doibase 10.1103/PhysRevLett.98.160502}
  {\bibfield  {journal} {\bibinfo  {journal} {Phys. Rev. Lett.}\ }\textbf
  {\bibinfo {volume} {98}},\ \bibinfo {pages} {160502} (\bibinfo {year}
  {2007})}\BibitemShut {NoStop}%
\bibitem [{\citenamefont {Bravyi}\ and\ \citenamefont
  {Kitaev}(1998)}]{Bravyi98}%
  \BibitemOpen
  \bibfield  {author} {\bibinfo {author} {\bibfnamefont {S.~B.}\ \bibnamefont
  {Bravyi}}\ and\ \bibinfo {author} {\bibfnamefont {A.~Y.}\ \bibnamefont
  {Kitaev}},\ }\href@noop {} {\bibfield  {journal} {\bibinfo  {journal} {arXiv
  preprint quant-ph/9811052}\ } (\bibinfo {year} {1998})}\BibitemShut {NoStop}%
\bibitem [{\citenamefont {Bravyi}\ and\ \citenamefont
  {K\"onig}(2013)}]{Bravyi13}%
  \BibitemOpen
  \bibfield  {author} {\bibinfo {author} {\bibfnamefont {S.}~\bibnamefont
  {Bravyi}}\ and\ \bibinfo {author} {\bibfnamefont {R.}~\bibnamefont
  {K\"onig}},\ }\href {\doibase 10.1103/PhysRevLett.110.170503} {\bibfield
  {journal} {\bibinfo  {journal} {Phys. Rev. Lett.}\ }\textbf {\bibinfo
  {volume} {110}},\ \bibinfo {pages} {170503} (\bibinfo {year}
  {2013})}\BibitemShut {NoStop}%
\bibitem [{\citenamefont {Aaronson}\ and\ \citenamefont
  {Gottesman}(2004)}]{Aaronson04}%
  \BibitemOpen
  \bibfield  {author} {\bibinfo {author} {\bibfnamefont {S.}~\bibnamefont
  {Aaronson}}\ and\ \bibinfo {author} {\bibfnamefont {D.}~\bibnamefont
  {Gottesman}},\ }\href {\doibase 10.1103/PhysRevA.70.052328} {\bibfield
  {journal} {\bibinfo  {journal} {Phys. Rev. A}\ }\textbf {\bibinfo {volume}
  {70}},\ \bibinfo {pages} {052328} (\bibinfo {year} {2004})}\BibitemShut
  {NoStop}%
\bibitem [{\citenamefont {Chamon}(2005)}]{Chamon05}%
  \BibitemOpen
  \bibfield  {author} {\bibinfo {author} {\bibfnamefont {C.}~\bibnamefont
  {Chamon}},\ }\href {\doibase 10.1103/PhysRevLett.94.040402} {\bibfield
  {journal} {\bibinfo  {journal} {Phys. Rev. Lett.}\ }\textbf {\bibinfo
  {volume} {94}},\ \bibinfo {pages} {040402} (\bibinfo {year}
  {2005})}\BibitemShut {NoStop}%
\bibitem [{\citenamefont {Bravyi}\ \emph {et~al.}(2011)\citenamefont {Bravyi},
  \citenamefont {Leemhuis},\ and\ \citenamefont {Terhal}}]{Bravyi11}%
  \BibitemOpen
  \bibfield  {author} {\bibinfo {author} {\bibfnamefont {S.}~\bibnamefont
  {Bravyi}}, \bibinfo {author} {\bibfnamefont {B.}~\bibnamefont {Leemhuis}}, \
  and\ \bibinfo {author} {\bibfnamefont {B.~M.}\ \bibnamefont {Terhal}},\
  }\href {\doibase https://doi.org/10.1016/j.aop.2010.11.002} {\bibfield
  {journal} {\bibinfo  {journal} {Annals of Physics}\ }\textbf {\bibinfo
  {volume} {326}},\ \bibinfo {pages} {839} (\bibinfo {year}
  {2011})}\BibitemShut {NoStop}%
\bibitem [{\citenamefont {Haah}(2011)}]{Haah11}%
  \BibitemOpen
  \bibfield  {author} {\bibinfo {author} {\bibfnamefont {J.}~\bibnamefont
  {Haah}},\ }\href {\doibase 10.1103/PhysRevA.83.042330} {\bibfield  {journal}
  {\bibinfo  {journal} {Phys. Rev. A}\ }\textbf {\bibinfo {volume} {83}},\
  \bibinfo {pages} {042330} (\bibinfo {year} {2011})}\BibitemShut {NoStop}%
\bibitem [{\citenamefont {Yoshida}(2013)}]{Yoshida13}%
  \BibitemOpen
  \bibfield  {author} {\bibinfo {author} {\bibfnamefont {B.}~\bibnamefont
  {Yoshida}},\ }\href {\doibase 10.1103/PhysRevB.88.125122} {\bibfield
  {journal} {\bibinfo  {journal} {Phys. Rev. B}\ }\textbf {\bibinfo {volume}
  {88}},\ \bibinfo {pages} {125122} (\bibinfo {year} {2013})}\BibitemShut
  {NoStop}%
\bibitem [{\citenamefont {Vijay}\ \emph {et~al.}(2015)\citenamefont {Vijay},
  \citenamefont {Haah},\ and\ \citenamefont {Fu}}]{Vijay15}%
  \BibitemOpen
  \bibfield  {author} {\bibinfo {author} {\bibfnamefont {S.}~\bibnamefont
  {Vijay}}, \bibinfo {author} {\bibfnamefont {J.}~\bibnamefont {Haah}}, \ and\
  \bibinfo {author} {\bibfnamefont {L.}~\bibnamefont {Fu}},\ }\href {\doibase
  10.1103/PhysRevB.92.235136} {\bibfield  {journal} {\bibinfo  {journal} {Phys.
  Rev. B}\ }\textbf {\bibinfo {volume} {92}},\ \bibinfo {pages} {235136}
  (\bibinfo {year} {2015})}\BibitemShut {NoStop}%
\bibitem [{\citenamefont {Vijay}\ \emph {et~al.}(2016)\citenamefont {Vijay},
  \citenamefont {Haah},\ and\ \citenamefont {Fu}}]{Vijay16}%
  \BibitemOpen
  \bibfield  {author} {\bibinfo {author} {\bibfnamefont {S.}~\bibnamefont
  {Vijay}}, \bibinfo {author} {\bibfnamefont {J.}~\bibnamefont {Haah}}, \ and\
  \bibinfo {author} {\bibfnamefont {L.}~\bibnamefont {Fu}},\ }\href {\doibase
  10.1103/PhysRevB.94.235157} {\bibfield  {journal} {\bibinfo  {journal} {Phys.
  Rev. B}\ }\textbf {\bibinfo {volume} {94}},\ \bibinfo {pages} {235157}
  (\bibinfo {year} {2016})}\BibitemShut {NoStop}%
\bibitem [{\citenamefont {Ma}\ \emph {et~al.}(2017)\citenamefont {Ma},
  \citenamefont {Lake}, \citenamefont {Chen},\ and\ \citenamefont
  {Hermele}}]{Ma17}%
  \BibitemOpen
  \bibfield  {author} {\bibinfo {author} {\bibfnamefont {H.}~\bibnamefont
  {Ma}}, \bibinfo {author} {\bibfnamefont {E.}~\bibnamefont {Lake}}, \bibinfo
  {author} {\bibfnamefont {X.}~\bibnamefont {Chen}}, \ and\ \bibinfo {author}
  {\bibfnamefont {M.}~\bibnamefont {Hermele}},\ }\href {\doibase
  10.1103/PhysRevB.95.245126} {\bibfield  {journal} {\bibinfo  {journal} {Phys.
  Rev. B}\ }\textbf {\bibinfo {volume} {95}},\ \bibinfo {pages} {245126}
  (\bibinfo {year} {2017})}\BibitemShut {NoStop}%
\bibitem [{\citenamefont {Vijay}\ and\ \citenamefont {Fu}(2017)}]{VijayNA}%
  \BibitemOpen
  \bibfield  {author} {\bibinfo {author} {\bibfnamefont {S.}~\bibnamefont
  {Vijay}}\ and\ \bibinfo {author} {\bibfnamefont {L.}~\bibnamefont {Fu}},\
  }\href@noop {} {\  (\bibinfo {year} {2017})},\ \Eprint
  {http://arxiv.org/abs/arXiv:1706.07070} {arXiv:1706.07070} \BibitemShut
  {NoStop}%
\bibitem [{\citenamefont {Nandkishore}\ and\ \citenamefont
  {Hermele}(2019)}]{Nandkishore19}%
  \BibitemOpen
  \bibfield  {author} {\bibinfo {author} {\bibfnamefont {R.~M.}\ \bibnamefont
  {Nandkishore}}\ and\ \bibinfo {author} {\bibfnamefont {M.}~\bibnamefont
  {Hermele}},\ }\href {\doibase 10.1146/annurev-conmatphys-031218-013604}
  {\bibfield  {journal} {\bibinfo  {journal} {Annual Review of Condensed Matter
  Physics}\ }\textbf {\bibinfo {volume} {10}},\ \bibinfo {pages} {295}
  (\bibinfo {year} {2019})}\BibitemShut {NoStop}%
\bibitem [{\citenamefont {Song}\ \emph {et~al.}(2019)\citenamefont {Song},
  \citenamefont {Prem}, \citenamefont {Huang},\ and\ \citenamefont
  {Martin-Delgado}}]{Song_Twisted}%
  \BibitemOpen
  \bibfield  {author} {\bibinfo {author} {\bibfnamefont {H.}~\bibnamefont
  {Song}}, \bibinfo {author} {\bibfnamefont {A.}~\bibnamefont {Prem}}, \bibinfo
  {author} {\bibfnamefont {S.-J.}\ \bibnamefont {Huang}}, \ and\ \bibinfo
  {author} {\bibfnamefont {M.~A.}\ \bibnamefont {Martin-Delgado}},\ }\href
  {\doibase 10.1103/PhysRevB.99.155118} {\bibfield  {journal} {\bibinfo
  {journal} {Phys. Rev. B}\ }\textbf {\bibinfo {volume} {99}},\ \bibinfo
  {pages} {155118} (\bibinfo {year} {2019})}\BibitemShut {NoStop}%
\bibitem [{\citenamefont {Prem}\ \emph {et~al.}(2019)\citenamefont {Prem},
  \citenamefont {Huang}, \citenamefont {Song},\ and\ \citenamefont
  {Hermele}}]{CageNet}%
  \BibitemOpen
  \bibfield  {author} {\bibinfo {author} {\bibfnamefont {A.}~\bibnamefont
  {Prem}}, \bibinfo {author} {\bibfnamefont {S.-J.}\ \bibnamefont {Huang}},
  \bibinfo {author} {\bibfnamefont {H.}~\bibnamefont {Song}}, \ and\ \bibinfo
  {author} {\bibfnamefont {M.}~\bibnamefont {Hermele}},\ }\href {\doibase
  10.1103/PhysRevX.9.021010} {\bibfield  {journal} {\bibinfo  {journal} {Phys.
  Rev. X}\ }\textbf {\bibinfo {volume} {9}},\ \bibinfo {pages} {021010}
  (\bibinfo {year} {2019})}\BibitemShut {NoStop}%
\bibitem [{\citenamefont {Wen}(2020)}]{WenDN}%
  \BibitemOpen
  \bibfield  {author} {\bibinfo {author} {\bibfnamefont {X.-G.}\ \bibnamefont
  {Wen}},\ }\href {\doibase 10.1103/PhysRevResearch.2.033300} {\bibfield
  {journal} {\bibinfo  {journal} {Phys. Rev. Research}\ }\textbf {\bibinfo
  {volume} {2}},\ \bibinfo {pages} {033300} (\bibinfo {year}
  {2020})}\BibitemShut {NoStop}%
\bibitem [{\citenamefont {Wang}(2020)}]{WangDN}%
  \BibitemOpen
  \bibfield  {author} {\bibinfo {author} {\bibfnamefont {J.}~\bibnamefont
  {Wang}},\ }\href@noop {} {\  (\bibinfo {year} {2020})},\ \Eprint
  {http://arxiv.org/abs/arXiv:2002.12932} {arXiv:2002.12932} \BibitemShut
  {NoStop}%
\bibitem [{\citenamefont {Aasen}\ \emph {et~al.}(2020)\citenamefont {Aasen},
  \citenamefont {Bulmash}, \citenamefont {Prem}, \citenamefont {Slagle},\ and\
  \citenamefont {Williamson}}]{AasenDN}%
  \BibitemOpen
  \bibfield  {author} {\bibinfo {author} {\bibfnamefont {D.}~\bibnamefont
  {Aasen}}, \bibinfo {author} {\bibfnamefont {D.}~\bibnamefont {Bulmash}},
  \bibinfo {author} {\bibfnamefont {A.}~\bibnamefont {Prem}}, \bibinfo {author}
  {\bibfnamefont {K.}~\bibnamefont {Slagle}}, \ and\ \bibinfo {author}
  {\bibfnamefont {D.~J.}\ \bibnamefont {Williamson}},\ }\href {\doibase
  10.1103/PhysRevResearch.2.043165} {\bibfield  {journal} {\bibinfo  {journal}
  {Phys. Rev. Research}\ }\textbf {\bibinfo {volume} {2}},\ \bibinfo {pages}
  {043165} (\bibinfo {year} {2020})}\BibitemShut {NoStop}%
\bibitem [{\citenamefont {Devakul}\ \emph {et~al.}(2018)\citenamefont
  {Devakul}, \citenamefont {Parameswaran},\ and\ \citenamefont
  {Sondhi}}]{Devakul18}%
  \BibitemOpen
  \bibfield  {author} {\bibinfo {author} {\bibfnamefont {T.}~\bibnamefont
  {Devakul}}, \bibinfo {author} {\bibfnamefont {S.~A.}\ \bibnamefont
  {Parameswaran}}, \ and\ \bibinfo {author} {\bibfnamefont {S.~L.}\
  \bibnamefont {Sondhi}},\ }\href {\doibase 10.1103/PhysRevB.97.041110}
  {\bibfield  {journal} {\bibinfo  {journal} {Phys. Rev. B}\ }\textbf {\bibinfo
  {volume} {97}},\ \bibinfo {pages} {041110(R)} (\bibinfo {year}
  {2018})}\BibitemShut {NoStop}%
\bibitem [{\citenamefont {M\"uhlhauser}\ \emph {et~al.}(2020)\citenamefont
  {M\"uhlhauser}, \citenamefont {Walther}, \citenamefont {Reiss},\ and\
  \citenamefont {Schmidt}}]{Muhlhauser20}%
  \BibitemOpen
  \bibfield  {author} {\bibinfo {author} {\bibfnamefont {M.}~\bibnamefont
  {M\"uhlhauser}}, \bibinfo {author} {\bibfnamefont {M.~R.}\ \bibnamefont
  {Walther}}, \bibinfo {author} {\bibfnamefont {D.~A.}\ \bibnamefont {Reiss}},
  \ and\ \bibinfo {author} {\bibfnamefont {K.~P.}\ \bibnamefont {Schmidt}},\
  }\href {\doibase 10.1103/PhysRevB.101.054426} {\bibfield  {journal} {\bibinfo
   {journal} {Phys. Rev. B}\ }\textbf {\bibinfo {volume} {101}},\ \bibinfo
  {pages} {054426} (\bibinfo {year} {2020})}\BibitemShut {NoStop}%
\bibitem [{\citenamefont {Zhou}\ \emph {et~al.}(2022)\citenamefont {Zhou},
  \citenamefont {Li}, \citenamefont {Yan}, \citenamefont {Ye},\ and\
  \citenamefont {Meng}}]{Zhou22}%
  \BibitemOpen
  \bibfield  {author} {\bibinfo {author} {\bibfnamefont {C.}~\bibnamefont
  {Zhou}}, \bibinfo {author} {\bibfnamefont {M.-Y.}\ \bibnamefont {Li}},
  \bibinfo {author} {\bibfnamefont {Z.}~\bibnamefont {Yan}}, \bibinfo {author}
  {\bibfnamefont {P.}~\bibnamefont {Ye}}, \ and\ \bibinfo {author}
  {\bibfnamefont {Z.~Y.}\ \bibnamefont {Meng}},\ }\href {\doibase
  10.1103/PhysRevResearch.4.033111} {\bibfield  {journal} {\bibinfo  {journal}
  {Phys. Rev. Research}\ }\textbf {\bibinfo {volume} {4}},\ \bibinfo {pages}
  {033111} (\bibinfo {year} {2022})}\BibitemShut {NoStop}%
\bibitem [{\citenamefont {Bravyi}\ and\ \citenamefont
  {Haah}(2013)}]{RGdecoder}%
  \BibitemOpen
  \bibfield  {author} {\bibinfo {author} {\bibfnamefont {S.}~\bibnamefont
  {Bravyi}}\ and\ \bibinfo {author} {\bibfnamefont {J.}~\bibnamefont {Haah}},\
  }\href {\doibase 10.1103/PhysRevLett.111.200501} {\bibfield  {journal}
  {\bibinfo  {journal} {Phys. Rev. Lett.}\ }\textbf {\bibinfo {volume} {111}},\
  \bibinfo {pages} {200501} (\bibinfo {year} {2013})}\BibitemShut {NoStop}%
\bibitem [{\citenamefont {Brown}\ and\ \citenamefont
  {Williamson}(2020)}]{Xcube_EC2019}%
  \BibitemOpen
  \bibfield  {author} {\bibinfo {author} {\bibfnamefont {B.~J.}\ \bibnamefont
  {Brown}}\ and\ \bibinfo {author} {\bibfnamefont {D.~J.}\ \bibnamefont
  {Williamson}},\ }\href {\doibase 10.1103/PhysRevResearch.2.013303} {\bibfield
   {journal} {\bibinfo  {journal} {Phys. Rev. Research}\ }\textbf {\bibinfo
  {volume} {2}},\ \bibinfo {pages} {013303} (\bibinfo {year}
  {2020})}\BibitemShut {NoStop}%
\bibitem [{\citenamefont {Verresen}\ \emph {et~al.}(2021)\citenamefont
  {Verresen}, \citenamefont {Tantivasadakarn},\ and\ \citenamefont
  {Vishwanath}}]{Verresen21}%
  \BibitemOpen
  \bibfield  {author} {\bibinfo {author} {\bibfnamefont {R.}~\bibnamefont
  {Verresen}}, \bibinfo {author} {\bibfnamefont {N.}~\bibnamefont
  {Tantivasadakarn}}, \ and\ \bibinfo {author} {\bibfnamefont {A.}~\bibnamefont
  {Vishwanath}},\ }\href@noop {} {\bibfield  {journal} {\bibinfo  {journal}
  {arXiv preprint arXiv:2112.03061}\ } (\bibinfo {year} {2021})}\BibitemShut
  {NoStop}%
\bibitem [{\citenamefont {Myerson-Jain}\ \emph {et~al.}(2022)\citenamefont
  {Myerson-Jain}, \citenamefont {Yan}, \citenamefont {Weld},\ and\
  \citenamefont {Xu}}]{Myerson22}%
  \BibitemOpen
  \bibfield  {author} {\bibinfo {author} {\bibfnamefont {N.~E.}\ \bibnamefont
  {Myerson-Jain}}, \bibinfo {author} {\bibfnamefont {S.}~\bibnamefont {Yan}},
  \bibinfo {author} {\bibfnamefont {D.}~\bibnamefont {Weld}}, \ and\ \bibinfo
  {author} {\bibfnamefont {C.}~\bibnamefont {Xu}},\ }\href {\doibase
  10.1103/PhysRevLett.128.017601} {\bibfield  {journal} {\bibinfo  {journal}
  {Phys. Rev. Lett.}\ }\textbf {\bibinfo {volume} {128}},\ \bibinfo {pages}
  {017601} (\bibinfo {year} {2022})}\BibitemShut {NoStop}%
\bibitem [{\citenamefont {You}\ and\ \citenamefont {von Oppen}(2019)}]{You19}%
  \BibitemOpen
  \bibfield  {author} {\bibinfo {author} {\bibfnamefont {Y.}~\bibnamefont
  {You}}\ and\ \bibinfo {author} {\bibfnamefont {F.}~\bibnamefont {von
  Oppen}},\ }\href {\doibase 10.1103/PhysRevResearch.1.013011} {\bibfield
  {journal} {\bibinfo  {journal} {Phys. Rev. Research}\ }\textbf {\bibinfo
  {volume} {1}},\ \bibinfo {pages} {013011} (\bibinfo {year}
  {2019})}\BibitemShut {NoStop}%
\bibitem [{\citenamefont {Katzgraber}\ \emph {et~al.}(2009)\citenamefont
  {Katzgraber}, \citenamefont {Bombin},\ and\ \citenamefont
  {Martin-Delgado}}]{Katzgraber09}%
  \BibitemOpen
  \bibfield  {author} {\bibinfo {author} {\bibfnamefont {H.~G.}\ \bibnamefont
  {Katzgraber}}, \bibinfo {author} {\bibfnamefont {H.}~\bibnamefont {Bombin}},
  \ and\ \bibinfo {author} {\bibfnamefont {M.~A.}\ \bibnamefont
  {Martin-Delgado}},\ }\href {\doibase 10.1103/PhysRevLett.103.090501}
  {\bibfield  {journal} {\bibinfo  {journal} {Phys. Rev. Lett.}\ }\textbf
  {\bibinfo {volume} {103}},\ \bibinfo {pages} {090501} (\bibinfo {year}
  {2009})}\BibitemShut {NoStop}%
\bibitem [{\citenamefont {Bombin}\ \emph {et~al.}(2012)\citenamefont {Bombin},
  \citenamefont {Andrist}, \citenamefont {Ohzeki}, \citenamefont {Katzgraber},\
  and\ \citenamefont {Martin-Delgado}}]{Bombin12}%
  \BibitemOpen
  \bibfield  {author} {\bibinfo {author} {\bibfnamefont {H.}~\bibnamefont
  {Bombin}}, \bibinfo {author} {\bibfnamefont {R.~S.}\ \bibnamefont {Andrist}},
  \bibinfo {author} {\bibfnamefont {M.}~\bibnamefont {Ohzeki}}, \bibinfo
  {author} {\bibfnamefont {H.~G.}\ \bibnamefont {Katzgraber}}, \ and\ \bibinfo
  {author} {\bibfnamefont {M.~A.}\ \bibnamefont {Martin-Delgado}},\ }\href
  {\doibase 10.1103/PhysRevX.2.021004} {\bibfield  {journal} {\bibinfo
  {journal} {Phys. Rev. X}\ }\textbf {\bibinfo {volume} {2}},\ \bibinfo {pages}
  {021004} (\bibinfo {year} {2012})}\BibitemShut {NoStop}%
\bibitem [{\citenamefont {Katzgraber}\ \emph {et~al.}(2010)\citenamefont
  {Katzgraber}, \citenamefont {Bombin}, \citenamefont {Andrist},\ and\
  \citenamefont {Martin-Delgado}}]{Katzgraber10}%
  \BibitemOpen
  \bibfield  {author} {\bibinfo {author} {\bibfnamefont {H.~G.}\ \bibnamefont
  {Katzgraber}}, \bibinfo {author} {\bibfnamefont {H.}~\bibnamefont {Bombin}},
  \bibinfo {author} {\bibfnamefont {R.~S.}\ \bibnamefont {Andrist}}, \ and\
  \bibinfo {author} {\bibfnamefont {M.~A.}\ \bibnamefont {Martin-Delgado}},\
  }\href {\doibase 10.1103/PhysRevA.81.012319} {\bibfield  {journal} {\bibinfo
  {journal} {Phys. Rev. A}\ }\textbf {\bibinfo {volume} {81}},\ \bibinfo
  {pages} {012319} (\bibinfo {year} {2010})}\BibitemShut {NoStop}%
\bibitem [{\citenamefont {Andrist}\ \emph {et~al.}(2011)\citenamefont
  {Andrist}, \citenamefont {Katzgraber}, \citenamefont {Bombin},\ and\
  \citenamefont {Martin-Delgado}}]{Andrist_2011}%
  \BibitemOpen
  \bibfield  {author} {\bibinfo {author} {\bibfnamefont {R.~S.}\ \bibnamefont
  {Andrist}}, \bibinfo {author} {\bibfnamefont {H.~G.}\ \bibnamefont
  {Katzgraber}}, \bibinfo {author} {\bibfnamefont {H.}~\bibnamefont {Bombin}},
  \ and\ \bibinfo {author} {\bibfnamefont {M.~A.}\ \bibnamefont
  {Martin-Delgado}},\ }\href {\doibase 10.1088/1367-2630/13/8/083006}
  {\bibfield  {journal} {\bibinfo  {journal} {New Journal of Physics}\ }\textbf
  {\bibinfo {volume} {13}},\ \bibinfo {pages} {083006} (\bibinfo {year}
  {2011})}\BibitemShut {NoStop}%
\bibitem [{\citenamefont {Andrist}\ \emph {et~al.}(2016)\citenamefont
  {Andrist}, \citenamefont {Katzgraber}, \citenamefont {Bombin},\ and\
  \citenamefont {Martin-Delgado}}]{Andrist16}%
  \BibitemOpen
  \bibfield  {author} {\bibinfo {author} {\bibfnamefont {R.~S.}\ \bibnamefont
  {Andrist}}, \bibinfo {author} {\bibfnamefont {H.~G.}\ \bibnamefont
  {Katzgraber}}, \bibinfo {author} {\bibfnamefont {H.}~\bibnamefont {Bombin}},
  \ and\ \bibinfo {author} {\bibfnamefont {M.~A.}\ \bibnamefont
  {Martin-Delgado}},\ }\href {\doibase 10.1103/PhysRevA.94.012318} {\bibfield
  {journal} {\bibinfo  {journal} {Phys. Rev. A}\ }\textbf {\bibinfo {volume}
  {94}},\ \bibinfo {pages} {012318} (\bibinfo {year} {2016})}\BibitemShut
  {NoStop}%
\bibitem [{\citenamefont {Kubica}\ \emph {et~al.}(2018)\citenamefont {Kubica},
  \citenamefont {Beverland}, \citenamefont {Brand\~ao}, \citenamefont
  {Preskill},\ and\ \citenamefont {Svore}}]{Kubica18}%
  \BibitemOpen
  \bibfield  {author} {\bibinfo {author} {\bibfnamefont {A.}~\bibnamefont
  {Kubica}}, \bibinfo {author} {\bibfnamefont {M.~E.}\ \bibnamefont
  {Beverland}}, \bibinfo {author} {\bibfnamefont {F.}~\bibnamefont
  {Brand\~ao}}, \bibinfo {author} {\bibfnamefont {J.}~\bibnamefont {Preskill}},
  \ and\ \bibinfo {author} {\bibfnamefont {K.~M.}\ \bibnamefont {Svore}},\
  }\href {\doibase 10.1103/PhysRevLett.120.180501} {\bibfield  {journal}
  {\bibinfo  {journal} {Phys. Rev. Lett.}\ }\textbf {\bibinfo {volume} {120}},\
  \bibinfo {pages} {180501} (\bibinfo {year} {2018})}\BibitemShut {NoStop}%
\bibitem [{\citenamefont {Viyuela}\ \emph {et~al.}(2019)\citenamefont
  {Viyuela}, \citenamefont {Vijay},\ and\ \citenamefont {Fu}}]{Viyuela19}%
  \BibitemOpen
  \bibfield  {author} {\bibinfo {author} {\bibfnamefont {O.}~\bibnamefont
  {Viyuela}}, \bibinfo {author} {\bibfnamefont {S.}~\bibnamefont {Vijay}}, \
  and\ \bibinfo {author} {\bibfnamefont {L.}~\bibnamefont {Fu}},\ }\href
  {\doibase 10.1103/PhysRevB.99.205114} {\bibfield  {journal} {\bibinfo
  {journal} {Phys. Rev. B}\ }\textbf {\bibinfo {volume} {99}},\ \bibinfo
  {pages} {205114} (\bibinfo {year} {2019})}\BibitemShut {NoStop}%
\bibitem [{\citenamefont {Vodola}\ \emph {et~al.}(2022)\citenamefont {Vodola},
  \citenamefont {Rispler}, \citenamefont {Kim},\ and\ \citenamefont
  {M{\"{u}}ller}}]{Vodola22}%
  \BibitemOpen
  \bibfield  {author} {\bibinfo {author} {\bibfnamefont {D.}~\bibnamefont
  {Vodola}}, \bibinfo {author} {\bibfnamefont {M.}~\bibnamefont {Rispler}},
  \bibinfo {author} {\bibfnamefont {S.}~\bibnamefont {Kim}}, \ and\ \bibinfo
  {author} {\bibfnamefont {M.}~\bibnamefont {M{\"{u}}ller}},\ }\href {\doibase
  10.22331/q-2022-01-05-618} {\bibfield  {journal} {\bibinfo  {journal}
  {{Quantum}}\ }\textbf {\bibinfo {volume} {6}},\ \bibinfo {pages} {618}
  (\bibinfo {year} {2022})}\BibitemShut {NoStop}%
\bibitem [{\citenamefont {Wang}\ \emph {et~al.}(2003)\citenamefont {Wang},
  \citenamefont {Harrington},\ and\ \citenamefont {Preskill}}]{Wang02}%
  \BibitemOpen
  \bibfield  {author} {\bibinfo {author} {\bibfnamefont {C.}~\bibnamefont
  {Wang}}, \bibinfo {author} {\bibfnamefont {J.}~\bibnamefont {Harrington}}, \
  and\ \bibinfo {author} {\bibfnamefont {J.}~\bibnamefont {Preskill}},\ }\href
  {\doibase https://doi.org/10.1016/S0003-4916(02)00019-2} {\bibfield
  {journal} {\bibinfo  {journal} {Annals of Physics}\ }\textbf {\bibinfo
  {volume} {303}},\ \bibinfo {pages} {31} (\bibinfo {year} {2003})}\BibitemShut
  {NoStop}%
\bibitem [{\citenamefont {Ohno}\ \emph {et~al.}(2004)\citenamefont {Ohno},
  \citenamefont {Arakawa}, \citenamefont {Ichinose},\ and\ \citenamefont
  {Matsui}}]{Ohno2004}%
  \BibitemOpen
  \bibfield  {author} {\bibinfo {author} {\bibfnamefont {T.}~\bibnamefont
  {Ohno}}, \bibinfo {author} {\bibfnamefont {G.}~\bibnamefont {Arakawa}},
  \bibinfo {author} {\bibfnamefont {I.}~\bibnamefont {Ichinose}}, \ and\
  \bibinfo {author} {\bibfnamefont {T.}~\bibnamefont {Matsui}},\ }\href
  {\doibase 10.1016/j.nuclphysb.2004.07.003} {\bibfield  {journal} {\bibinfo
  {journal} {Nuclear Physics B}\ }\textbf {\bibinfo {volume} {697}},\ \bibinfo
  {pages} {462} (\bibinfo {year} {2004})}\BibitemShut {NoStop}%
\bibitem [{SM()}]{SM}%
  \BibitemOpen
  \href@noop {} {}\bibinfo {note} {{See Supplemental Material for the
  Kramers-Wannier duality of CSS code, more details of the error correction
  process, the proof for the absence of glass order along Nishimori line, and
  details of numerical simulations, which contains additional Refs.~\cite{Wu76,
  BookNishimori, Lee90, Lee91, Jin12, Baxter73, Mueller14, Challa86,
  Katzgraber06, BookEfron}.}}\BibitemShut {Stop}%
\bibitem [{\citenamefont {Calderbank}\ \emph {et~al.}(1997)\citenamefont
  {Calderbank}, \citenamefont {Rains}, \citenamefont {Shor},\ and\
  \citenamefont {Sloane}}]{CSS}%
  \BibitemOpen
  \bibfield  {author} {\bibinfo {author} {\bibfnamefont {A.~R.}\ \bibnamefont
  {Calderbank}}, \bibinfo {author} {\bibfnamefont {E.~M.}\ \bibnamefont
  {Rains}}, \bibinfo {author} {\bibfnamefont {P.~W.}\ \bibnamefont {Shor}}, \
  and\ \bibinfo {author} {\bibfnamefont {N.~J.~A.}\ \bibnamefont {Sloane}},\
  }\href {\doibase 10.1103/PhysRevLett.78.405} {\bibfield  {journal} {\bibinfo
  {journal} {Phys. Rev. Lett.}\ }\textbf {\bibinfo {volume} {78}},\ \bibinfo
  {pages} {405} (\bibinfo {year} {1997})}\BibitemShut {NoStop}%
\bibitem [{\citenamefont {Ditzian}\ \emph {et~al.}(1980)\citenamefont
  {Ditzian}, \citenamefont {Banavar}, \citenamefont {Grest},\ and\
  \citenamefont {Kadanoff}}]{Kadanoff80}%
  \BibitemOpen
  \bibfield  {author} {\bibinfo {author} {\bibfnamefont {R.~V.}\ \bibnamefont
  {Ditzian}}, \bibinfo {author} {\bibfnamefont {J.~R.}\ \bibnamefont
  {Banavar}}, \bibinfo {author} {\bibfnamefont {G.~S.}\ \bibnamefont {Grest}},
  \ and\ \bibinfo {author} {\bibfnamefont {L.~P.}\ \bibnamefont {Kadanoff}},\
  }\href {\doibase 10.1103/PhysRevB.22.2542} {\bibfield  {journal} {\bibinfo
  {journal} {Phys. Rev. B}\ }\textbf {\bibinfo {volume} {22}},\ \bibinfo
  {pages} {2542} (\bibinfo {year} {1980})}\BibitemShut {NoStop}%
\bibitem [{\citenamefont {Johnston}\ and\ \citenamefont
  {Ranasinghe}(2011)}]{Johnston11}%
  \BibitemOpen
  \bibfield  {author} {\bibinfo {author} {\bibfnamefont {D.~A.}\ \bibnamefont
  {Johnston}}\ and\ \bibinfo {author} {\bibfnamefont {R.~P. K. C.~M.}\
  \bibnamefont {Ranasinghe}},\ }\href {\doibase 10.1088/1751-8113/44/29/295004}
  {\bibfield  {journal} {\bibinfo  {journal} {Journal of Physics A:
  Mathematical and Theoretical}\ }\textbf {\bibinfo {volume} {44}},\ \bibinfo
  {pages} {295004} (\bibinfo {year} {2011})}\BibitemShut {NoStop}%
\bibitem [{\citenamefont {Johnston}\ \emph {et~al.}(2017)\citenamefont
  {Johnston}, \citenamefont {Mueller},\ and\ \citenamefont
  {Janke}}]{Johnston17}%
  \BibitemOpen
  \bibfield  {author} {\bibinfo {author} {\bibfnamefont {D.~A.}\ \bibnamefont
  {Johnston}}, \bibinfo {author} {\bibfnamefont {M.}~\bibnamefont {Mueller}}, \
  and\ \bibinfo {author} {\bibfnamefont {W.}~\bibnamefont {Janke}},\ }\href
  {\doibase 10.1140/epjst/e2016-60329-4} {\bibfield  {journal} {\bibinfo
  {journal} {The European Physical Journal Special Topics}\ }\textbf {\bibinfo
  {volume} {226}},\ \bibinfo {pages} {749} (\bibinfo {year}
  {2017})}\BibitemShut {NoStop}%
\bibitem [{\citenamefont {Imry}\ and\ \citenamefont {Ma}(1975)}]{Imry75}%
  \BibitemOpen
  \bibfield  {author} {\bibinfo {author} {\bibfnamefont {Y.}~\bibnamefont
  {Imry}}\ and\ \bibinfo {author} {\bibfnamefont {S.-k.}\ \bibnamefont {Ma}},\
  }\href {\doibase 10.1103/PhysRevLett.35.1399} {\bibfield  {journal} {\bibinfo
   {journal} {Phys. Rev. Lett.}\ }\textbf {\bibinfo {volume} {35}},\ \bibinfo
  {pages} {1399} (\bibinfo {year} {1975})}\BibitemShut {NoStop}%
\bibitem [{\citenamefont {Imry}\ and\ \citenamefont {Wortis}(1979)}]{Imry79}%
  \BibitemOpen
  \bibfield  {author} {\bibinfo {author} {\bibfnamefont {Y.}~\bibnamefont
  {Imry}}\ and\ \bibinfo {author} {\bibfnamefont {M.}~\bibnamefont {Wortis}},\
  }\href {\doibase 10.1103/PhysRevB.19.3580} {\bibfield  {journal} {\bibinfo
  {journal} {Phys. Rev. B}\ }\textbf {\bibinfo {volume} {19}},\ \bibinfo
  {pages} {3580} (\bibinfo {year} {1979})}\BibitemShut {NoStop}%
\bibitem [{\citenamefont {Janke}(2008)}]{Janke08}%
  \BibitemOpen
  \bibfield  {author} {\bibinfo {author} {\bibfnamefont {W.}~\bibnamefont
  {Janke}},\ }\enquote {\bibinfo {title} {Monte carlo methods in classical
  statistical physics},}\ in\ \href {\doibase 10.1007/978-3-540-74686-7_4}
  {\emph {\bibinfo {booktitle} {Computational Many-Particle Physics}}},\
  \bibinfo {editor} {edited by\ \bibinfo {editor} {\bibfnamefont
  {H.}~\bibnamefont {Fehske}}, \bibinfo {editor} {\bibfnamefont
  {R.}~\bibnamefont {Schneider}}, \ and\ \bibinfo {editor} {\bibfnamefont
  {A.}~\bibnamefont {Wei{\ss}e}}}\ (\bibinfo  {publisher} {Springer Berlin
  Heidelberg},\ \bibinfo {address} {Berlin, Heidelberg},\ \bibinfo {year}
  {2008})\ pp.\ \bibinfo {pages} {79--140}\BibitemShut {NoStop}%
\bibitem [{\citenamefont {{Gilbert}}(1952)}]{Gilbert52}%
  \BibitemOpen
  \bibfield  {author} {\bibinfo {author} {\bibfnamefont {E.~N.}\ \bibnamefont
  {{Gilbert}}},\ }\href {\doibase 10.1002/j.1538-7305.1952.tb01393.x}
  {\bibfield  {journal} {\bibinfo  {journal} {The Bell System Technical
  Journal}\ }\textbf {\bibinfo {volume} {31}},\ \bibinfo {pages} {504}
  (\bibinfo {year} {1952})}\BibitemShut {NoStop}%
\bibitem [{\citenamefont {Varshamov}(1957)}]{Varshamov57}%
  \BibitemOpen
  \bibfield  {author} {\bibinfo {author} {\bibfnamefont {R.~R.}\ \bibnamefont
  {Varshamov}},\ }\href@noop {} {\bibfield  {journal} {\bibinfo  {journal}
  {Docklady Akad. Nauk, S.S.S.R.}\ }\textbf {\bibinfo {volume} {117}},\
  \bibinfo {pages} {739} (\bibinfo {year} {1957})}\BibitemShut {NoStop}%
\bibitem [{\citenamefont {Calderbank}\ and\ \citenamefont
  {Shor}(1996)}]{Calderbank96}%
  \BibitemOpen
  \bibfield  {author} {\bibinfo {author} {\bibfnamefont {A.~R.}\ \bibnamefont
  {Calderbank}}\ and\ \bibinfo {author} {\bibfnamefont {P.~W.}\ \bibnamefont
  {Shor}},\ }\href {\doibase 10.1103/PhysRevA.54.1098} {\bibfield  {journal}
  {\bibinfo  {journal} {Phys. Rev. A}\ }\textbf {\bibinfo {volume} {54}},\
  \bibinfo {pages} {1098} (\bibinfo {year} {1996})}\BibitemShut {NoStop}%
\bibitem [{\citenamefont {Nishimori}(2007)}]{Nishimori2007}%
  \BibitemOpen
  \bibfield  {author} {\bibinfo {author} {\bibfnamefont {H.}~\bibnamefont
  {Nishimori}},\ }\href {\doibase 10.1007/s10955-006-9156-1} {\bibfield
  {journal} {\bibinfo  {journal} {Journal of Statistical Physics}\ }\textbf
  {\bibinfo {volume} {126}},\ \bibinfo {pages} {977} (\bibinfo {year}
  {2007})},\ \bibinfo {note} {where a different convention is used with $1-p$
  (instead of $p$) denoting the probability for antiferromagnetic coupling, and
  see also the references therein.}\BibitemShut {Stop}%
\bibitem [{\citenamefont {Gaenko}\ \emph {et~al.}(2017)\citenamefont {Gaenko},
  \citenamefont {Antipov}, \citenamefont {Carcassi}, \citenamefont {Chen},
  \citenamefont {Chen}, \citenamefont {Dong}, \citenamefont {Gamper},
  \citenamefont {Gukelberger}, \citenamefont {Igarashi}, \citenamefont
  {Iskakov}, \citenamefont {K{\"o}nz}, \citenamefont {LeBlanc}, \citenamefont
  {Levy}, \citenamefont {Ma}, \citenamefont {Paki}, \citenamefont {Shinaoka},
  \citenamefont {Todo}, \citenamefont {Troyer},\ and\ \citenamefont
  {Gull}}]{Gaenko17}%
  \BibitemOpen
  \bibfield  {author} {\bibinfo {author} {\bibfnamefont {A.}~\bibnamefont
  {Gaenko}}, \bibinfo {author} {\bibfnamefont {A.}~\bibnamefont {Antipov}},
  \bibinfo {author} {\bibfnamefont {G.}~\bibnamefont {Carcassi}}, \bibinfo
  {author} {\bibfnamefont {T.}~\bibnamefont {Chen}}, \bibinfo {author}
  {\bibfnamefont {X.}~\bibnamefont {Chen}}, \bibinfo {author} {\bibfnamefont
  {Q.}~\bibnamefont {Dong}}, \bibinfo {author} {\bibfnamefont {L.}~\bibnamefont
  {Gamper}}, \bibinfo {author} {\bibfnamefont {J.}~\bibnamefont {Gukelberger}},
  \bibinfo {author} {\bibfnamefont {R.}~\bibnamefont {Igarashi}}, \bibinfo
  {author} {\bibfnamefont {S.}~\bibnamefont {Iskakov}}, \bibinfo {author}
  {\bibfnamefont {M.}~\bibnamefont {K{\"o}nz}}, \bibinfo {author}
  {\bibfnamefont {J.}~\bibnamefont {LeBlanc}}, \bibinfo {author} {\bibfnamefont
  {R.}~\bibnamefont {Levy}}, \bibinfo {author} {\bibfnamefont {P.}~\bibnamefont
  {Ma}}, \bibinfo {author} {\bibfnamefont {J.}~\bibnamefont {Paki}}, \bibinfo
  {author} {\bibfnamefont {H.}~\bibnamefont {Shinaoka}}, \bibinfo {author}
  {\bibfnamefont {S.}~\bibnamefont {Todo}}, \bibinfo {author} {\bibfnamefont
  {M.}~\bibnamefont {Troyer}}, \ and\ \bibinfo {author} {\bibfnamefont
  {E.}~\bibnamefont {Gull}},\ }\href {\doibase 10.1016/j.cpc.2016.12.009}
  {\bibfield  {journal} {\bibinfo  {journal} {Comput. Phys. Commun.}\ }\textbf
  {\bibinfo {volume} {213}},\ \bibinfo {pages} {235} (\bibinfo {year}
  {2017})}\BibitemShut {NoStop}%
\bibitem [{\citenamefont {Greitemann}\ \emph {et~al.}(2019)\citenamefont
  {Greitemann}, \citenamefont {Liu},\ and\ \citenamefont
  {Pollet}}]{Greitemann19}%
  \BibitemOpen
  \bibfield  {author} {\bibinfo {author} {\bibfnamefont {J.}~\bibnamefont
  {Greitemann}}, \bibinfo {author} {\bibfnamefont {K.}~\bibnamefont {Liu}}, \
  and\ \bibinfo {author} {\bibfnamefont {L.}~\bibnamefont {Pollet}},\ }\href
  {\doibase 10.1103/PhysRevB.99.060404} {\bibfield  {journal} {\bibinfo
  {journal} {Phys. Rev. B}\ }\textbf {\bibinfo {volume} {99}},\ \bibinfo
  {pages} {060404(R)} (\bibinfo {year} {2019})}\BibitemShut {NoStop}%
\bibitem [{\citenamefont {Liu}\ \emph {et~al.}(2019)\citenamefont {Liu},
  \citenamefont {Greitemann},\ and\ \citenamefont {Pollet}}]{Liu19}%
  \BibitemOpen
  \bibfield  {author} {\bibinfo {author} {\bibfnamefont {K.}~\bibnamefont
  {Liu}}, \bibinfo {author} {\bibfnamefont {J.}~\bibnamefont {Greitemann}}, \
  and\ \bibinfo {author} {\bibfnamefont {L.}~\bibnamefont {Pollet}},\ }\href
  {\doibase 10.1103/PhysRevB.99.104410} {\bibfield  {journal} {\bibinfo
  {journal} {Phys. Rev. B}\ }\textbf {\bibinfo {volume} {99}},\ \bibinfo
  {pages} {104410} (\bibinfo {year} {2019})}\BibitemShut {NoStop}%
\bibitem [{dat()}]{data}%
  \BibitemOpen
  \href@noop {} {}\bibinfo {howpublished}
  {\url{https://github.com/KeLiu-04/Xcube_data}}\BibitemShut {NoStop}%
\bibitem [{\citenamefont {Wu}\ and\ \citenamefont {Wang}(1976)}]{Wu76}%
  \BibitemOpen
  \bibfield  {author} {\bibinfo {author} {\bibfnamefont {F.~Y.}\ \bibnamefont
  {Wu}}\ and\ \bibinfo {author} {\bibfnamefont {Y.~K.}\ \bibnamefont {Wang}},\
  }\href {\doibase 10.1063/1.522914} {\bibfield  {journal} {\bibinfo  {journal}
  {Journal of Mathematical Physics}\ }\textbf {\bibinfo {volume} {17}},\
  \bibinfo {pages} {439} (\bibinfo {year} {1976})}\BibitemShut {NoStop}%
\bibitem [{\citenamefont {Nishimori}(2001)}]{BookNishimori}%
  \BibitemOpen
  \bibfield  {author} {\bibinfo {author} {\bibfnamefont {H.}~\bibnamefont
  {Nishimori}},\ }\href@noop {} {\emph {\bibinfo {title} {Statistical Physics
  of Spin Glasses and Information Processing: an Introduction}}}\ (\bibinfo
  {publisher} {Oxford University Press},\ \bibinfo {address} {Oxford; New
  York},\ \bibinfo {year} {2001})\BibitemShut {NoStop}%
\bibitem [{\citenamefont {Lee}\ and\ \citenamefont {Kosterlitz}(1990)}]{Lee90}%
  \BibitemOpen
  \bibfield  {author} {\bibinfo {author} {\bibfnamefont {J.}~\bibnamefont
  {Lee}}\ and\ \bibinfo {author} {\bibfnamefont {J.~M.}\ \bibnamefont
  {Kosterlitz}},\ }\href {\doibase 10.1103/PhysRevLett.65.137} {\bibfield
  {journal} {\bibinfo  {journal} {Phys. Rev. Lett.}\ }\textbf {\bibinfo
  {volume} {65}},\ \bibinfo {pages} {137} (\bibinfo {year} {1990})}\BibitemShut
  {NoStop}%
\bibitem [{\citenamefont {Lee}\ and\ \citenamefont {Kosterlitz}(1991)}]{Lee91}%
  \BibitemOpen
  \bibfield  {author} {\bibinfo {author} {\bibfnamefont {J.}~\bibnamefont
  {Lee}}\ and\ \bibinfo {author} {\bibfnamefont {J.~M.}\ \bibnamefont
  {Kosterlitz}},\ }\href {\doibase 10.1103/PhysRevB.43.3265} {\bibfield
  {journal} {\bibinfo  {journal} {Phys. Rev. B}\ }\textbf {\bibinfo {volume}
  {43}},\ \bibinfo {pages} {3265} (\bibinfo {year} {1991})}\BibitemShut
  {NoStop}%
\bibitem [{\citenamefont {Jin}\ \emph {et~al.}(2012)\citenamefont {Jin},
  \citenamefont {Sen},\ and\ \citenamefont {Sandvik}}]{Jin12}%
  \BibitemOpen
  \bibfield  {author} {\bibinfo {author} {\bibfnamefont {S.}~\bibnamefont
  {Jin}}, \bibinfo {author} {\bibfnamefont {A.}~\bibnamefont {Sen}}, \ and\
  \bibinfo {author} {\bibfnamefont {A.~W.}\ \bibnamefont {Sandvik}},\ }\href
  {\doibase 10.1103/PhysRevLett.108.045702} {\bibfield  {journal} {\bibinfo
  {journal} {Phys. Rev. Lett.}\ }\textbf {\bibinfo {volume} {108}},\ \bibinfo
  {pages} {045702} (\bibinfo {year} {2012})}\BibitemShut {NoStop}%
\bibitem [{\citenamefont {Baxter}(1973)}]{Baxter73}%
  \BibitemOpen
  \bibfield  {author} {\bibinfo {author} {\bibfnamefont {R.~J.}\ \bibnamefont
  {Baxter}},\ }\href {\doibase 10.1088/0022-3719/6/23/005} {\bibfield
  {journal} {\bibinfo  {journal} {Journal of Physics C: Solid State Physics}\
  }\textbf {\bibinfo {volume} {6}},\ \bibinfo {pages} {L445} (\bibinfo {year}
  {1973})}\BibitemShut {NoStop}%
\bibitem [{\citenamefont {Mueller}\ \emph {et~al.}(2014)\citenamefont
  {Mueller}, \citenamefont {Janke},\ and\ \citenamefont
  {Johnston}}]{Mueller14}%
  \BibitemOpen
  \bibfield  {author} {\bibinfo {author} {\bibfnamefont {M.}~\bibnamefont
  {Mueller}}, \bibinfo {author} {\bibfnamefont {W.}~\bibnamefont {Janke}}, \
  and\ \bibinfo {author} {\bibfnamefont {D.~A.}\ \bibnamefont {Johnston}},\
  }\href {\doibase 10.1103/PhysRevLett.112.200601} {\bibfield  {journal}
  {\bibinfo  {journal} {Phys. Rev. Lett.}\ }\textbf {\bibinfo {volume} {112}},\
  \bibinfo {pages} {200601} (\bibinfo {year} {2014})}\BibitemShut {NoStop}%
\bibitem [{\citenamefont {Challa}\ \emph {et~al.}(1986)\citenamefont {Challa},
  \citenamefont {Landau},\ and\ \citenamefont {Binder}}]{Challa86}%
  \BibitemOpen
  \bibfield  {author} {\bibinfo {author} {\bibfnamefont {M.~S.~S.}\
  \bibnamefont {Challa}}, \bibinfo {author} {\bibfnamefont {D.~P.}\
  \bibnamefont {Landau}}, \ and\ \bibinfo {author} {\bibfnamefont
  {K.}~\bibnamefont {Binder}},\ }\href {\doibase 10.1103/PhysRevB.34.1841}
  {\bibfield  {journal} {\bibinfo  {journal} {Phys. Rev. B}\ }\textbf {\bibinfo
  {volume} {34}},\ \bibinfo {pages} {1841} (\bibinfo {year}
  {1986})}\BibitemShut {NoStop}%
\bibitem [{\citenamefont {Katzgraber}\ \emph {et~al.}(2006)\citenamefont
  {Katzgraber}, \citenamefont {Trebst}, \citenamefont {Huse},\ and\
  \citenamefont {Troyer}}]{Katzgraber06}%
  \BibitemOpen
  \bibfield  {author} {\bibinfo {author} {\bibfnamefont {H.~G.}\ \bibnamefont
  {Katzgraber}}, \bibinfo {author} {\bibfnamefont {S.}~\bibnamefont {Trebst}},
  \bibinfo {author} {\bibfnamefont {D.~A.}\ \bibnamefont {Huse}}, \ and\
  \bibinfo {author} {\bibfnamefont {M.}~\bibnamefont {Troyer}},\ }\href
  {\doibase 10.1088/1742-5468/2006/03/p03018} {\bibfield  {journal} {\bibinfo
  {journal} {Journal of Statistical Mechanics: Theory and Experiment}\ }\textbf
  {\bibinfo {volume} {2006}},\ \bibinfo {pages} {P03018} (\bibinfo {year}
  {2006})}\BibitemShut {NoStop}%
\bibitem [{\citenamefont {Efron}\ and\ \citenamefont
  {Tibshirani}(1994)}]{BookEfron}%
  \BibitemOpen
  \bibfield  {author} {\bibinfo {author} {\bibfnamefont {B.}~\bibnamefont
  {Efron}}\ and\ \bibinfo {author} {\bibfnamefont {R.~J.}\ \bibnamefont
  {Tibshirani}},\ }\href@noop {} {\emph {\bibinfo {title} {An introduction to
  the bootstrap}}}\ (\bibinfo  {publisher} {CRC press},\ \bibinfo {year}
  {1994})\BibitemShut {NoStop}%
\end{thebibliography}%
%
%
%

\onecolumngrid
\clearpage
\makeatletter
\begin{center}
  \textbf{\large --- Supplementary Materials ---\\[0.5em]Optimal Thresholds for Fracton Codes and Random Spin Models \\ with Subsystem Symmetry}\\[1em]

Hao Song, Janik Sch\"onmeier-Kromer, Ke Liu, Oscar Viyuela, Lode Pollet, and M. A. Martin-Delgado
  \thispagestyle{titlepage}
\end{center}
\setcounter{equation}{0}
\setcounter{figure}{0}
\setcounter{table}{0}
\setcounter{page}{1}
\setcounter{section}{0}
\renewcommand{\theequation}{S\arabic{equation}}
\renewcommand{\thefigure}{S\arabic{figure}}
\renewcommand{\thetable}{S\arabic{table}}
\renewcommand{\thesection}{S.\Roman{section}}

\section{Kramers-Wannier duality for CSS code}
Here we show that the 3D random plaquette Ising (RPI) model $H_{\eta}^{\mathcal{A}}$ and the 3D random anisotropically coupled Ashkin-Teller (RACAT) model $H_{\eta}^{\mathcal{B}}$ can be related by a  Kramers-Wannier duality.
Our discussion is not restricted to specific models and can apply to the analysis of general fracton and topological Calderbank-Shor-Steane (CSS) codes.

\subsection{Exact Kramers-Wannier duality for disorder-free models}

The duality between $H_{\eta}^{\mathcal{A}}$ and $H_{\eta}^{\mathcal{B}}$ is exact in the disorder-free ($p=0$) limit. 
In this limit, the relevant error equivalence classes are the trivial ones ($\eta \equiv 0$), and the partition function $\mathcal{Z}_{\eta}^{\mathcal{A}}$ modeling $X$-errors reduces to
\begin{align}
\mathcal{Z}_{0}^{\mathcal{A}}\left(\beta\right) 
= \sum_{f\in\mathbb{Z}_{2}^{\mathcal{A}}} e^{\beta\sum_{\ell\in\mathcal{Q}}\left(-1\right)^{\partial_{A}f\left(\ell\right)}}
= \sum_{f\in\mathbb{Z}_{2}^{\mathcal{A}}} W_{\beta}\left(\partial_A f\right),
\end{align}
where $f \equiv \{f(c)\}_{c\in\mathcal{A}} \in \mathbb{Z}_2^{\mathcal{A}}$ labels the configuration of type-$A$ stabilizer generators, 
$\partial_{A} f \in \mathbb{Z}_2^{\mathcal{Q}}$ specifies the corresponding qubit configuration, and 
$W_{\beta}(\xi)\coloneqq \prod_{\ell} e^{\beta (-1)^{\xi(\ell)}}$ denotes the Boltzmann weight for a general qubit configuration $\xi \in \mathbb{Z}_2^\mathcal{Q}$.

The Kramers-Wannier duality can be viewed as a Fourier transform~\cite{Wu76}. The dual of $W_{\beta}$ can be expressed as
\begin{align}\label{eq:dual_W0}
\widetilde{W}_{\beta}\left(\zeta\right) 
&\coloneqq 2^{-\frac{\left|\mathcal{Q}\right|}{2}}\sum_{\xi \in\mathbb{Z}_{2}^{\mathcal{Q}}}W_{\beta}\left(\xi \right)e^{-i\pi\left\langle \zeta,\xi \right\rangle} \nonumber \\
& =\prod_{\ell\in\mathcal{Q}} \sqrt{\sinh 2\beta} \, e^{\tilde{\beta}\left(-1\right)^{\zeta(\ell)}} 
= \left(\sinh 2\beta\right)^{\frac{\left|\mathcal{Q}\right|}{2}}W_{\tilde{\beta}}\left(\zeta\right),
\end{align}
in terms of a dual inverse temperature $\tilde{\beta}$ specified by
\begin{align} \label{eq:beta_dual}
\sinh\left(2\beta\right)\sinh(2\tilde{\beta})=1,
\end{align}
where $\zeta \in \mathbb{Z}_2^{\mathcal{Q}}$ is the conjugate variable of $\xi$, and 
$\left\langle \zeta,\xi \right\rangle \coloneqq \sum_{\ell\in\mathcal{Q}} \zeta\left(\ell\right) \xi \left(\ell\right)$ 
denotes an inner product.

The Fourier transform rewrites $\mathcal{Z}_{0}^{\mathcal{A}}$ as 
\begin{align}\label{eq:Z_A}
\mathcal{Z}_{0}^{\mathcal{A}}\left(\beta\right) 
= 2^{-\frac{\left|\mathcal{Q}\right|}{2}}  \sum_{f\in\mathbb{Z}_{2}^{\mathcal{A}}}\sum_{\zeta\in\mathbb{Z}_{2}^{\mathcal{Q}}} \widetilde{W}_{\beta}\left(\zeta\right)e^{i\pi\left\langle \zeta,\partial_{A}f\right\rangle}.
\end{align}
Using the identities
$\left\langle \zeta,\partial_{A}f\right\rangle =\left\langle \partial_{A}^{\dagger}\zeta,f\right\rangle $
and $\sum_{f_\in{\mathbb{Z}^{\mathcal{A}}_2}}e^{i\pi\left\langle \partial_{A}^{\dagger}\zeta,f\right\rangle } = 2^{\left|\mathcal{A}\right|}\delta\left(\partial_{A}^{\dagger}\zeta\right)$,
one finds that only those $\zeta \in \ker\partial^{\dagger}_A$ contribute.
Moreover, as in the thermodynamical limit the free energy density is independent of the boundary conditions, we can choose an open boundary condition such that $\ker \partial^{\dagger}_A = {\rm im} \, \partial_B \simeq \mathbb{Z}_{2}^{\mathcal{B}}/\ker\partial_{B}$.
Thus, Eq.~\eqref{eq:Z_A} becomes
\begin{align}
\mathcal{Z}_{0}^{\mathcal{A}}\left(\beta\right) 
= \frac{2^{\left|\mathcal{A}\right|-\frac{\left|\mathcal{Q}\right|}{2}}}{\left|\ker\partial_B\right|}
\sum_{g\in\mathbb{Z}_{2}^{\mathcal{B}}} \widetilde{W}_{\beta}\left(\partial_{B}g\right), \label{eq:dual_Z1} 
\end{align}
with $g$ labelling configurations of type-$B$ stabilizer generators, and $\partial_B g = \zeta \in {\rm im} \, \partial_B$.

Therefore, the Kramers-Wannier duality for CSS codes in the disorder-free limit can be established as 
\begin{align}
\mathcal{Z}_{0}^{\mathcal{A}}\left(\beta\right) 
= \frac{2^{\left|\mathcal{A}\right|-\frac{\left|\mathcal{Q}\right|}{2}}}{\left|\ker\partial_B\right|}
\left(\sinh 2\beta\right)^{\frac{\left|\mathcal{Q}\right|}{2}}
\sum_{g\in\mathbb{Z}_{2}^{\mathcal{B}}} W_{\tilde{\beta}}\left(\partial_{B}g\right)
= \frac{2^{\left|\mathcal{A}\right|-\frac{\left|\mathcal{Q}\right|}{2}}}{\left|\ker\partial_B\right|} \left(\sinh 2\beta\right)^{\frac{\left|\mathcal{Q}\right|}{2}}\mathcal{Z}_{0}^{\mathcal{B}}(\tilde{\beta}). \label{eq:dual_Z} 
\end{align}
As the X-cube model is a CSS code, the duality between the disorder-free limit of the RPI model and the RACAT model, namely, $H_{\eta = 0}^{\mathcal{A}}$ and $H_{\eta=0}^{\mathcal{B}}$, then follows immediately. 

\subsection{Approximate duality between the optimal bit-flip and phase-flip error thresholds}

In the presence of disorder ($p > 0$), there is no exact duality between the pair of error models $H_{\eta}^{\mathcal{A}}$ and $H_{\eta}^{\mathcal{B}}$. 
Nevertheless, the current work for error models with subsystem symmetries and the previous studies for models with global or local symmetries~\cite{Nishimori2007, Katzgraber09, Kubica18} suggest that the optimal error thresholds $p_c^X$ and $p_c^Z$ satisfy an approximate duality relation
\begin{align}\label{eq:approx_dual}
H\left(p_{c}^{X}\right)+H\left(p_{c}^{Z}\right) \approx 1,
\end{align}
where $H\left(p\right)\coloneqq-p\log_{2}\left(p\right)-\left(1-p\right)\log_{2}\left(1-p\right)$ is the Shannon entropy.
Below, we show that this approximate duality may be understood by a replica analysis. 

Consider a disorder average of the partition function $\mathcal{Z}_{\eta}^{\mathcal{A}}$ over $n$ replicas,
\begin{align}\label{eq:Zn_A}
\mathcal{Z}_{n}^{\mathcal{A}} 
\coloneqq\left[\prod_{j=1}^{n} \sum_{f_j \in\mathbb{Z}_{2}^{\mathcal{A}}}e^{\beta\sum_{\ell\in\mathcal{Q}}\left(-1\right)^{\eta(\ell)+\partial_{A}f_j\left(\ell\right)}}\right]
= \sum_{\mathbf{f} \in \left(\mathbb{Z}_{2}^{\mathcal{A}}\right)^n}\prod_{\ell\in\mathcal{Q}}\left[\prod_{j=1}^{n} e^{\beta\left(-1\right)^{\eta(\ell)+\partial_{A}f_j\left(\ell\right)}}\right],
\end{align}
where $f_j \in\mathbb{Z}_{2}^{\mathcal{A}}$ labels configurations of type-$A$ stabilizer generators for the $j$-th replica, and $\mathbf{f} \coloneqq (f_1, f_2, \cdots, f_n) \in \left(\mathbb{Z}_{2}^{\mathcal{A}}\right)^n$.
At the error rate $p$, the coupling coefficient $(-1)^{\eta(\ell)}$ equals to $\pm 1$ with the probability $1-p$ and $p$, respectively.
Thus, the disorder-averaged Boltzmann factor associated with each edge can be expressed as
\begin{align}\label{eq:Wp}
w_{p,\beta}\left(|\mathbf{h}|\right) 
=\left[\prod_{j=1}^{n} e^{\beta\left(-1\right)^{\eta(\ell)+h_j}}\right]
=\left(1-p\right)e^{\left(n-2|\mathbf{h}|\right)\beta}+ pe^{-\left(n-2|\mathbf{h}|\right)\beta},
\end{align}
where $h_j = \partial_A f_j(\ell) \in \{0, 1\}$, 
$\mathbf{h} \coloneqq \left(h_{1},h_{2},\cdots,h_{n}\right)\in\mathbb{Z}_{2}^{n}$,
and $|\mathbf{h}|$ counts the number of the $h_j = 1$ components.

As in the disorder-free case, we can define a dual Boltzmann factor
\begin{align}\label{eq:dual_Wp}
	\widetilde{w}_{p,\beta}\left(|\mathbf{t}|\right) & \coloneqq2^{-\frac{n}{2}}\sum_{\mathbf{h}\in\mathbb{Z}_{2}^{n}}w_{p,\beta}\left(|\mathbf{h}|\right)e^{-i\pi\left\langle \mathbf{h},\mathbf{t}\right\rangle }=\begin{cases}
2^{-\frac{n}{2}}\left(e^{\beta}+e^{-\beta}\right)^{n-|\mathbf{t}|}\left(e^{\beta}-e^{-\beta}\right)^{|\mathbf{t}|} & |\mathbf{t}|\text{ even}\\
2^{-\frac{n}{2}}\left(1-2p\right)\left(e^{\beta}+e^{-\beta}\right)^{n-|\mathbf{t}|}\left(e^{\beta}-e^{-\beta}\right)^{|\mathbf{t}|} & |\mathbf{t}|\text{ odd}
\end{cases},
\end{align}
with $\mathbf{t} \coloneqq (t_1, t_2, \cdots, t_n) \in\mathbb{Z}_{2}^{n}$ being the conjugate variable of $\mathbf{h}$, and $|\mathbf{t}|$ denoting the number of $t_j = 1$.
Analogous to Eq.~\eqref{eq:dual_Z1}, the disorder-averaged partition functions $\mathcal{Z}_{n}^{\mathcal{A}}$ and $\mathcal{Z}_{n}^{\mathcal{B}}$ can be related as
\begin{align}\label{eq:Zn_dual}
\mathcal{Z}_{n}^{\mathcal{A}}\left(w_{p,\beta}\left(|\mathbf{h}|\right) \right) 
= \frac{2^{n\left|\mathcal{A}\right|-\frac{n}{2}\left|\mathcal{Q}\right|}}{|\ker\partial_B|^n} \mathcal{Z}_{n}^{\mathcal{B}}\left(\widetilde{w}_{p,\beta}\left(|\mathbf{t}|\right)\right),
\end{align}
where $\mathcal{Z}_{n}^{\mathcal{A}}$ and $\mathcal{Z}_{n}^{\mathcal{B}}$ are viewed as functions of $n+1$ Boltzmann factors labelled by $|\mathbf{h}|$ or $|\mathbf{t}| = 0, 1, \cdots, n$.

With the principle factors
$w_{\rm P}(p, \beta) \coloneqq \max w_{p,\beta}\left(|\mathbf{h}|\right) = w_{p,\beta}(0)$ 
and $\widetilde{w}_{\rm P}(p, \beta) \coloneqq
\max\widetilde{w}_{p,\beta}\left(|\mathbf{t}|\right)=\widetilde{w}_{p,\beta}\left(0\right)$ factored out, 
Eq.~\eqref{eq:Zn_dual} becomes
\begin{align}\label{eq:dual_AB}
w_{\rm P}^{\left|\mathcal{Q}\right|} (p, \beta) \, \mathcal{Z}_{n}^{\mathcal{A}}\left(\mathbf{u}(p, \beta)\right) 
= \frac{2^{n\left|\mathcal{A}\right|-\frac{n}{2}\left|\mathcal{Q}\right|}}{|\ker\partial_B|^n} \, \widetilde{w}_{\rm P}^{\left|\mathcal{Q}\right|}(p, \beta) \, \mathcal{Z}_{n}^{\mathcal{B}}\left(\widetilde{\mathbf{u}}(p, \beta)\right),
\end{align}
where $\mathbf{u}(p, \beta) \coloneqq \frac{1}{w_{\rm P}(p, \beta)} (w_{p,\beta}(0), w_{p,\beta}(1), \cdots, w_{p,\beta}(n))$
and $\widetilde{\mathbf{u}}(p, \beta) \coloneqq \frac{1}{\widetilde{w}_{\rm P}(p, \beta)} (\widetilde{w}_{p,\beta}(0), \widetilde{w}_{p,\beta}(1), \cdots, \widetilde{w}_{p,\beta}(n))$
are the normalized Boltzmann factors.

Analogously, we also have
\begin{align}\label{eq:dual_BA}
w_{\rm P}^{\left|\mathcal{Q}\right|} (p, \beta) \, \mathcal{Z}_{n}^{\mathcal{B}}\left(\mathbf{u}(p, \beta) \right) 
= \frac{2^{n\left|\mathcal{B}\right|-\frac{n}{2}\left|\mathcal{Q}\right|}}{|\ker\partial_A|^n} \, \widetilde{w}_{\rm P}^{\left|\mathcal{Q}\right|}(p, \beta) \, \mathcal{Z}_{n}^{\mathcal{A}}\left(\widetilde{\mathbf{u}}(p, \beta) \right).
\end{align}

Assume that, for both $H_{\eta}^{\mathcal{A}}$ and $H_{\eta}^{\mathcal{B}}$, there are only one ordered phase and one disordered phase separated by a single phase transition at $\left(p_{c}^{X},\beta\left(p_{c}^{X}\right)\right)$ and $\left(p_{c}^{Z},\beta\left(p_{c}^{Z}\right)\right)$, respectively, on the Nishimori line 
$\beta=\beta(p)$, where
\begin{align}\label{eq:NL_SM}
	\beta\left(p\right) \coloneqq
	-\frac{1}{2}\ln\frac{p}{1-p}.
\end{align} 
The phase transition of $\mathcal{Z}_{\eta}^{\mathcal{A}}$ ($\mathcal{Z}_{\eta}^{\mathcal{B}}$) will then occur along the path of normalized Boltzmann factors $\mathbf{u}(p,\beta(p))$ at $p_c^X$ ($p_c^Z$), 
and also along the dual path $\widetilde{\mathbf{u}}(p,\beta(p))$ at $p_c^Z$ ($p_c^X$), respectively, using the relations in Eqs.~\eqref{eq:dual_BA} and \eqref{eq:dual_AB}.
As the phase transition is unique, one may expect $\mathbf{u}\left(p_{c}^{X},\beta\left(p_{c}^{X}\right)\right)\approx\widetilde{\mathbf{u}}\left(p_{c}^{Z},\beta\left(p_{c}^{Z}\right)\right)$
and $\mathbf{u}\left(p_{c}^{Z},\beta\left(p_{c}^{Z}\right)\right)\approx\widetilde{\mathbf{u}}\left(p_{c}^{X},\beta\left(p_{c}^{X}\right)\right)$.
Hence, by multiplying Eq.~\eqref{eq:dual_AB} and Eq.~\eqref{eq:dual_BA}, 
and using the identity $2^{\left|\mathcal{A}\right|+\left|\mathcal{B}\right|-\left|\mathcal{Q}\right|}=\left|\ker\partial_{A}\right|\left|\ker\partial_{B}\right|$,
we obtain the following relation between the principle Boltzmann factors
$w_{{\rm P}}$ and $\widetilde{w}_{{\rm P}}$,
\begin{align}
w_{{\rm P}}\left(p_{c}^{X},\beta\left(p_{c}^{X}\right)\right)w_{{\rm P}}\left(p_{c}^{Z},\beta\left(p_{c}^{Z}\right)\right)\approx\widetilde{w}_{{\rm P}}\left(p_{c}^{X},\beta\left(p_{c}^{X}\right)\right)\widetilde{w}_{{\rm P}}\left(p_{c}^{Z},\beta\left(p_{c}^{Z}\right)\right).\label{eq:dual_priciple_factor}
\end{align}

Moreover, given the expressions of the Boltzmann factors in  Eq.~\eqref{eq:Wp} and Eq.~\eqref{eq:dual_Wp}, one has
\begin{align} 
\ln\frac{w_{\rm P}\left(p,\beta\left(p\right)\right)}{\widetilde{w}_{\rm P}\left(p,\beta\left(p\right)\right)} 
=
\ln\frac{w_{p,\beta(p)}(0)}{\widetilde{w}_{p,\beta(p)}(0)} 
= \ln\left\{ 2^{\frac{n}{2}}\left[\left(1-p\right)^{n+1}+p^{n+1}\right]\right\} 
= n\ln2\left[\frac{1}{2}  - H\left(p\right)\right]+\mathcal{O}\left(n^{2}\right)
\end{align}
in the replica limit $n\rightarrow 0$ and along the Nishimori line Eq.~\eqref{eq:NL_SM}.

Therefore, the approximate duality relation Eq.~\eqref{eq:approx_dual} can be established, namely,
\begin{align}
\lim_{n\rightarrow0}\frac{1}{n\ln2}\ln\frac{w_{\rm P}\left(p_{c}^{X},\beta\left(p_{c}^{X}\right)\right) w_{\rm P}\left(p_{c}^{Z},\beta\left(p_{c}^{Z}\right)\right)}{\widetilde{w}_{\rm P}\left(p_{c}^{X},\beta\left(p_{c}^{X}\right)\right) \widetilde{w}_{\rm P}\left(p_{c}^{Z},\beta\left(p_{c}^{Z}\right)\right)}
= 1 - H\left(p_{c}^{X}\right) - H\left(p_{c}^{Z}\right) \approx 0.
\end{align}

\section{Error probability and correction}\label{sec:correction}

To be able to recover encoded quantum information from qubit errors, one needs to identify the error equivalence class unambiguously.
For clarity, we take the detection of bit-flip ($X$) errors as an example, whereas phase-flips ($Z$) errors can be analyzed in the same way.

Let $\rho=\left|\psi\right\rangle \left\langle \psi\right|$ be the density matrix of a generic state in the X-cube code space and $\mathrm{pr}(\eta)$ denote the probability of an $X$ error configuration $\eta$.
The error affected state $\rho_1$ is given by
\begin{align}\label{eq:rho_01}
\rho_{1} &=\sum_{\eta} \mathrm{pr} (\eta) X_{\eta} \,\rho \, X_{\eta} \\
	&= \sum_{\left[\eta\right]_X} \mathrm{pr} (\left[\eta\right]_X) X_{\eta} \,\rho\,X_{\eta},
\end{align}
where in the second line we have grouped error configurations into equivalence class $\left[\eta\right]_X \coloneqq \eta+\mathrm{im\,}\partial_{A}$.

We relabel error equivalence classes by syndromes and logical operators.
This has a practical relevance: Error syndromes are local measurements of stabilizer generators and cannot distinguish topological operators, such as $X$ logical operators which are strings winding around the lattice. 
Namely, a set of distinct $X$ error equivalence classes 
$\left[\eta_\sigma + \lambda \right]_{X}\coloneqq \eta_\sigma + \lambda + \mathrm{im\,}\partial_{A}$
will lead to the identical syndrome $\sigma$, where $\lambda$ labels inequivalent logical operators.
Such a surjective relation between error equivalence classes and error syndromes roots in the nature of a topological code.  
Hence, $\rho_1$ can be expressed as
\begin{equation}\label{eq:rho_1}
\rho_{1}=\sum_{\sigma,\lambda}\mathrm{pr}\left(\left[\eta_{\sigma} +\lambda\right]_{X}\right)X_{\eta_{\sigma}+\lambda}\,\rho\,X_{\eta_{\sigma}+\lambda}.
\end{equation}

Without loss of generality, we use $\left[\eta^*_{\sigma}\right]_{X}$ with $\lambda = 0$ to represent the most probable error class, such that
$\mathrm{pr} (\left[\eta^*_\sigma\right]_X) \geq \mathrm{pr} (\left[\eta^*_\sigma + \lambda \right]_X)$, $\forall \lambda$.
Then, if we clean the error syndrome by $X_{\eta^*_{\sigma}}$, the quantum state becomes
\begin{align}\label{eq:rho_12}
\rho_{2} = \sum_{\sigma}\mathrm{pr}\left(\left[\eta^*_{\sigma}\right]_{X}\right)  \rho
+ \sum_{\sigma} \sum_{\lambda \neq 0} \mathrm{pr}\left(\left[\eta^*_{\sigma} +\lambda\right]_{X}\right) X_{\lambda}\,\rho\,X_{\lambda},
\end{align}
To ensure the recovery of the initial state, namely, $\rho_2 \rightarrow \rho$, we shall have $\mathrm{pr}\left(\left[\eta^*_{\sigma} +\lambda\right]_{X}\right) / \mathrm{pr}\left(\left[\eta^*_{\sigma}\right]_{X}\right) \rightarrow 0$ for $\lambda \neq 0$ and large enough code blocks.
This is indeed possible if the code is in its ordered phase and, given our established mapping between the X-cube code and statistical-mechanical models, it can be demonstrated by a procedure originally developed for surface codes~\cite{Dennis02}.

It is intuitive to rewrite the Nishimori line in Eq.~\eqref{eq:NL_SM} as
\begin{equation}\label{eq:Nishimori2}
e^{-\frac{2}{T}}=\frac{p}{1-p}.
\end{equation}
The probability of each error configuration $\eta$ compatible with $\sigma$ then resembles a Boltzmann weight,  
\begin{equation}\label{eq:pr_SM}
\mathrm{pr}\left(\eta_\sigma;p\right) \propto\left(\frac{p}{1-p}\right)^{\sum_{\ell}\eta_\sigma\left(\ell\right)} 
= e^{-\frac{2}{T} \sum_{\ell}\eta_\sigma\left(\ell\right)}.
\end{equation}
Accordingly, the relative probability between two $X$ error equivalence classes relates to the partition function of the RPI model as (reproduced from the main text for convenience) 
\begin{align}\label{eq:deltaF_SM}
\frac{\text{pr}\left(\left[\eta^*_\sigma + \lambda\right]_{X};p\right)}{\text{pr}\left(\left[\eta^*_\sigma\right]_{X};p\right)}
= \frac{\mathcal{Z}_{\eta^*_\sigma+\lambda}^{\mathcal{A}}\left(\beta\right)}{\mathcal{Z}_{\eta^*_\sigma}^{\mathcal{A}}\left(\beta\right)}
=e^{-\beta\delta\mathcal{F}_{\eta^*_\sigma,\lambda}^{\mathcal{A}}}.
\end{align}
Therefore, the relative probability can be estimated by the free energy cost $\delta\mathcal{F}_{\eta^*_\sigma,\lambda \neq 0}^{\mathcal{A}}$ of a topological defect (generalization of domain walls) in $H_{\eta}^{\mathcal{A}}$. 
In the ordered phase of $H_{\eta}^{\mathcal{A}}$, such defects are suppressed exponentially.
In the disordered phase, topological defects costs zero energy, hence all error classes $\left[\eta_\sigma + \lambda \right]_{X}$ have equal probability and a most probable error class does not exist.

\section{Absence of glass order along Nishimori line \label{noGlass}}
Whether there exists a spin glass order can crucially affect the error thresholds of a CSS code.
In particular, in establishing the approximate duality relation Eq.~\eqref{eq:approx_dual}, the assumption of a unique phase transition also implies spin glass phases are irrelevant. 
Below, we prove that there is indeed no spin glass order along Nishimori line neither for the RPI model nor the RACAT model, by generalizing Nishimori's argument which was originally conceived for the Edwards-Anderson model~\cite{BookNishimori}.

We first focus on the RPI model, whose Hamiltonian can be expressed as
\begin{align}
H_{\eta}^{\mathcal{A}}(\tau; S)
= -\sum_{\ell \in \mathcal{Q}}\left(-1\right)^{\eta\left(\ell\right)}\prod_{c\in\partial_{A}^{\dagger}\ell}S_{c}
=-\sum_{\ell\in\mathcal{Q}}\tau_{\ell}\Gamma_{\ell}(S),
\end{align}
where $S \coloneqq \left\{S_{c}\right\} \in\{\pm 1\}^{\mathcal{A}}$ 
and $\tau \coloneqq \{\tau_{\ell}=\left(-1\right)^{\eta\left(\ell\right)}\} \in \mathbb{Z}_{2}^{\mathcal{Q}}$ 
label the configurations of spins and coupling coefficients, respectively,
and $\Gamma_{\ell}(S) \coloneqq \prod_{c\in\partial_{A}^{\dagger}\ell}S_{c}$ denotes the product of $S_c$ spins on the cubes sharing edge $\ell$.

The Hamiltonian $H_{\eta}^{\mathcal{A}}(\tau; S)$ has a $\mathbb{Z}_2$ gauge symmetry
\begin{align}\label{eq:gauge_transformation}
S \rightarrow \sigma S = \{\sigma_c S_c\}, \quad
\tau \rightarrow \Gamma\left(\sigma\right)\tau 
= \{\Gamma_{\ell}\left(\sigma\right)\tau_{\ell} \},
\quad \forall \sigma = \{\sigma_c\} \in \{\pm 1\}^{\mathcal{A}},
\end{align} 
where $\Gamma_{\ell}(\sigma) \coloneqq \prod_{c\in\partial_{A}^{\dagger}\ell}\sigma_{c}$ analogous to $\Gamma_{\ell}(S)$.
Hence, for any finite subset $\xi \subset \mathcal{A}$,  the thermal average of $S_{\xi}\coloneqq\prod_{c\in\xi}S_{c}$ satisfies 
\begin{align}\label{eq:S_xi_relation}
\left\langle S_{\xi}\right\rangle_{\tau,\beta}
\coloneqq \frac{\sum_{S}S_{\xi} e^{-\beta H^{\mathcal{A}}_\eta(\tau; S)}}{\sum_{S}e^{-\beta H^{\mathcal{A}}_\eta(\tau; S)}}
=\sigma_{\xi}\left\langle S_{\xi}\right\rangle_{\Gamma(\sigma)\tau,\beta}
\end{align} with $\sigma_{\xi}\coloneqq\prod_{c\in\xi}\sigma_{c}$. 
Consequently, at the error rate $p$, the disorder average of $\left\langle S_{\xi}\right\rangle_{\tau,\beta}$ satisfies 
\begin{align}\label{eq:S_xi_d}
\left[\left\langle S_{\xi}\right\rangle _{\tau,\beta}\right]_p 
& \coloneqq\sum_{\tau}\mathrm{pr}\left(\tau;p\right)\left\langle S_{\xi}\right\rangle _{\tau,\beta}
=\frac{1}{2^{\left|\mathcal{A}\right|}}\sum_{\tau}\sum_{\sigma}\mathrm{pr}\left(\tau;p\right)\sigma_{\xi}\left\langle S_{\xi}\right\rangle _{\Gamma\left(\sigma\right)\tau,\beta}.
\end{align}
Notice that the sum over $\tau$ is unchanged if we reindex $\tau \rightarrow \Gamma(\sigma)\tau$, namely,
\begin{align}\label{eq:sum_tau_id}
\sum_{\tau}\mathrm{pr}\left(\tau;p\right)\sigma_{\xi}\left\langle S_{\xi}\right\rangle _{\Gamma\left(\sigma\right)\tau,\beta}
= \sum_{\tau}\mathrm{pr}\left(\Gamma\left(\sigma\right)\tau;p\right)\sigma_{\xi}\left\langle S_{\xi}\right\rangle_{\tau,\beta}.
\end{align}
Moreover, as $p \propto e^{\beta(p)}$ and $1-p \propto e^{-\beta(p)}$ in terms of the auxiliary inverse temperature $\beta(p) \coloneqq -\frac{1}{2} \ln \frac{p}{1-p}$, one has
\begin{align}\label{eq:S_avN}
\frac{\sum_{\sigma}\mathrm{pr}\left(\Gamma\left(\sigma\right)\tau;p\right)\sigma_{\xi}}{\sum_{\sigma}\mathrm{pr}\left(\Gamma\left(\sigma\right)\tau;p\right)}
=\frac{\sum_{S}\mathrm{pr}\left(\Gamma\left(S\right)\tau;p\right)S_{\xi}}{\sum_{S}\mathrm{pr}\left(\Gamma\left(S\right)\tau;p\right)}
=\left\langle S_{\xi}\right\rangle _{\tau,\beta(p)}.
\end{align}
By Eq.~\eqref{eq:sum_tau_id} and Eq.~\eqref{eq:S_avN}, the disorder average $[\left\langle S_{\xi}\right\rangle_{\tau,\beta}]_p$ becomes 
\begin{align}
\left[\left\langle S_{\xi}\right\rangle_{\tau,\beta}\right]_p 
& =\frac{1}{2^{\left|\mathcal{A}\right|}}\sum_{\tau}\sum_{\sigma}\mathrm{pr}\left(\Gamma\left(\sigma\right)\tau;p\right)\left\langle S_{\xi}\right\rangle _{\tau,\beta(p)}\left\langle S_{\xi}\right\rangle _{\tau,\beta}.
\end{align}
Further, using Eq.~\eqref{eq:S_xi_relation} again, and noticing $\sigma_{\xi}^2=1$, and reindexing  $ \Gamma(\sigma)\tau \rightarrow \tau$, one realizes an identity
\begin{align}
\left[\left\langle S_{\xi}\right\rangle_{\tau,\beta}\right]_p 
& =\frac{1}{2^{\left|\mathcal{A}\right|}}\sum_{\tau}\sum_{\sigma}\mathrm{pr}\left(\Gamma\left(\sigma\right)\tau;p\right)
\left\langle S_{\xi}\right\rangle _{\Gamma(\sigma)\tau,\beta(p)}
\left\langle S_{\xi}\right\rangle _{\Gamma(\sigma)\tau,\beta} \nonumber \\
& = \sum_{\tau}\mathrm{pr}\left(\tau;p\right)\left\langle S_{\xi}\right\rangle _{\tau,\beta(p)}\left\langle
S_{\xi}\right\rangle _{\tau,\beta} 
\eqqcolon \left[\left\langle S_{\xi}\right\rangle _{\tau,\beta(p)}\left\langle S_{\xi}\right\rangle _{\tau,\beta}\right]_p.
\end{align}
In particular, on the Nishimori line $\beta=\beta(p)$, we have
\begin{align}\label{eq:NishimoriA}
\left[\left\langle S_{\xi}\right\rangle_{\tau,\beta(p)}\right]_p
= \left[\left\langle S_{\xi}\right\rangle_{\tau,\beta(p)}^{2}\right]_p,
\end{align}
which indicates an equivalence between a normal order paramter and a spin-glass (SG) order parameter.
Specifically, if we choose $S_\xi$ to be the RPI correlation function, the identity Eq.~\eqref{eq:NishimoriA} becomes 
$G^{\mathcal{A}}\left(\mathbf{r}\right) = G_{\mathrm{SG}}^{\mathcal{A}}\left(\mathbf{r}\right)$, with
\begin{align}
G^{\mathcal{A}}\left(\mathbf{r}\right) 
& \coloneqq \frac{1}{L^3} \sum_c \left[\left\langle S_{c}S_{c+\hat{z}}S_{c+\left(\mathbf{r},0\right)}S_{c+\left(\mathbf{r},1\right)}\right\rangle _{\tau,\beta(p)}\right]_p,\label{eq:GA_SM}\\
G_{\mathrm{SG}}^{\mathcal{A}}\left(\mathbf{r}\right) 
& \coloneqq \frac{1}{L^3} \sum_c \left[\left\langle S_{c}S_{c+\hat{z}}S_{c+\left(\mathbf{r},0\right)}S_{c+\left(\mathbf{r},1\right)}\right\rangle _{\tau,\beta(p)}^{2}\right]_p \label{eq:GA_SG_SM}.
\end{align}
In the ordered phase of the RPI model, both $G^{\mathcal{A}}\left(\mathbf{r}\right)$ and $G_{\mathrm{SG}}^{\mathcal{A}}\left(\mathbf{r}\right)$ will develop a finite value in the limit $\left|\mathbf{r}\right| \rightarrow \infty$, while they vanish in the disordered phase.
A spin glass phase would require
$\lim_{\left|\mathbf{r}\right|\rightarrow\infty}G^{\mathcal{A}}\left(\mathbf{r}\right) = 0$ 
but $\lim_{\left|\mathbf{r}\right|\rightarrow\infty}G_{\mathrm{SG}}^{\mathcal{A}}\left(\mathbf{r}\right) \neq 0$,
which is nevertheless precluded by Eq.~\eqref{eq:NishimoriA}.

Similarly, for the RACAT correlator, we can also derive an identity $G^{\mathcal{B}}\left(\mathbf{r}\right) = G_{\mathrm{SG}}^{\mathcal{B}}\left(\mathbf{r}\right)$, where
\begin{align}
	G^{\mathcal{B}}\left(\mathbf{r}\right) 
	&\coloneqq \frac{1}{L^3} \sum_{v} \left[\left\langle S_{v}^{\mu}S_{v+r\hat{\mu}}^{\mu}\right\rangle_{\tau,\beta(p)}\right]_p, \\
	G_{\mathrm{SG}}^{\mathcal{B}}\left(\mathbf{r}\right) 
	&\coloneqq \frac{1}{L^3} \sum_{v} \left[\left\langle S_{v}^{\mu}S_{v+r\hat{\mu}}^{\mu}\right\rangle _{\tau,\beta(p)}^{2}\right]_p.
\end{align}
Hence, a spin glass order is also absent in the RACAT model along the Nishimori line.

\begin{figure*}[t]
\centering
	\includegraphics{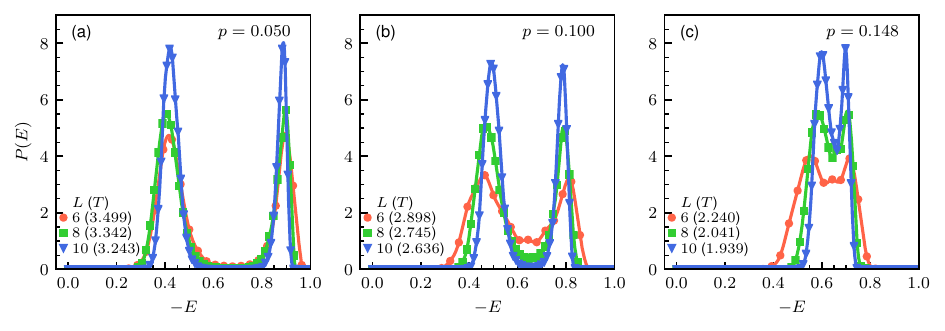}
	\caption{The energy histograms $P(E)$ for the 3D RPI model at different error rates. For small $p$ values, the double peaks in (a) and (b) grow sharper with increasing $L$, as expected for first-order phase transitions. When approaching the threshold value $p_c^X \simeq 0.152(4)$, the two peaks do not evolve towards separated $\delta$-functions as the weight of the valley retains a finite value with increasing $L$. Therefore, the features in (c) most likely correspond to finite-size effects of a continuous the phase transition.}
	\label{fig:PB_PIM}
\end{figure*}

\begin{figure*}[t]
\centering
	\includegraphics{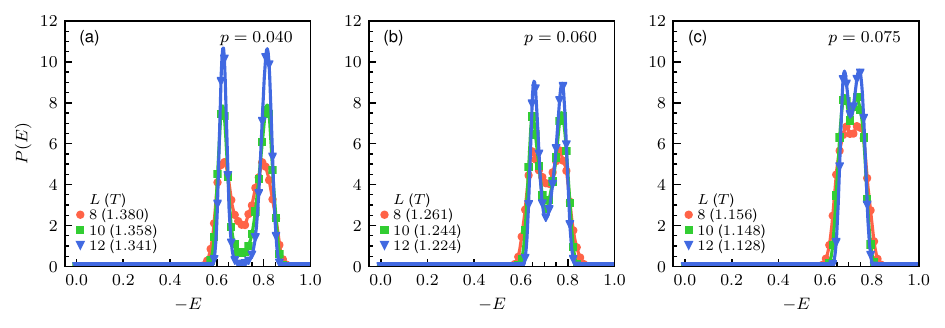}
	\caption{The histograms $P(E)$ for the 3D RACAT model at different error rates. As in the case of the RPI model, the system experiences discontinuous phase transitions in the low $p$ regime as shown in (a) and (b). The quenched disorder weakens the discontinuity as the distance $\Delta E$ between the two $P(E)$ peaks shrinks upon increasing $p$. The non-divergent peaks in (c) suggest finite-size effects of a second-order phase transition.}
	\label{fig:PB_ATM}
\end{figure*}

\section{Detecting phase transitions} \label{sec:transition}
The stability of the X-cube code against local $X$ and $Z$ errors can be related to the order-disorder phase transitions of the RPI and RACAT model, respectively.
Here we describe in detail how these phase transitions and the optimal error thresholds $p^X_c$ and $p^Z_c$ are determined.

\subsection{Observables and order of phase transitions}
For both models and each $p$ value, we compute the energy histogram $P(E)$, the specific heat $C_v$, the susceptibility $\chi$, and the second-moment correlation length $\xi_L$,
\begin{align}
	& P(E, \beta, L) = \left[\left\langle \delta(E - E^{\prime})\right\rangle  \right], \\
	& C_v(\beta, L) = \frac{L^3}{T^2}\left[ \left\langle E^{2}\right\rangle - \left\langle E \right\rangle^2 \right], \\
	& \chi(\beta, L) = \frac{L^3}{T}\left[ \left\langle Q^{2}\right\rangle - \left\langle Q \right\rangle^2 \right], \\
	& \xi_L(\beta, L) = \frac{1}{2\sin\left(\lvert\mathbf{k}_{\min}\rvert/2\right)}\left(\frac{\tilde{G}\left(\mathbf{0}\right)}{\tilde{G}\left(\mathbf{k}_{\min}\right)}-1\right)^{1/2}.
\end{align}
Here $E$ is the energy density of a state, $Q$ measures the value of the order parameters $Q^{\mathcal{A}}$ or $Q^{\mathcal{B}}$, and $\langle . \rangle$ and $[.]$ represent the thermal and disorder average, respectively.
The definition of $\xi_L$ is reproduced from the main text for convenience.

We find the order of the phase transitions by examining the behaviors of the energy histogram.
For continuous phase transitions, $P(E)$ typically shows a single peak for all temperatures.
When a phase transition is discontinuous, $P(E)$ features a double-peak structure on large enough system sizes and at temperatures near the transition, reflecting the phase separation of two metastable phases~\cite{Lee90}.
In addition, with increasing system sizes, the double peaks shall grow sharper and evolve towards separated $\delta$ functions.
This is what we observed in the low $p$ regions of the RPI model $H_{\eta}^{\mathcal{A}}$ and the RACAT model $H_{\eta}^{\mathcal{B}}$, as shown in Figure~\ref{fig:PB_PIM} and Figure~\ref{fig:PB_ATM}.

The distance $\Delta E$ between the double $P(E)$ peaks reflects the latent heat~\cite{Lee91}, which shrinks upon increasing $p$ as the quenched disorder weakens first-order phase transitions. 
Close to the error thresholds $p_c^X \simeq 0.152(4)$ and $p_c^Z \simeq 0.075(2)$, although $P(E)$ remains showing two peaks [Figure~\ref{fig:PB_PIM}(c) and Figure~\ref{fig:PB_ATM}(c)], the weight of the valley between them does not evolve towards zero when increasing $L$.
This implies the double peaks will not evolve into separate $\delta$-functions in the infinite size limit, in contrast to the case of a first-order phase transition.
Such a non-diverging behavior is also observed as a finite-size effect in simulations of the 2D $4$-state Potts model~\cite{Jin12}, where the phase transition is analytically known to be continuous~\cite{Baxter73}.

Therefore, these data suggest that, in the finite but low $p$ regions, the phase transitions of $H_{\eta}^{\mathcal{A}}$ and $H_{\eta}^{\mathcal{B}}$ are discontinuous similar as in the disorder-free limits~\cite{Mueller14, Johnston17}.
Nevertheless, in both cases, the first-order transition line terminates at a critical point, where the $p$ value coincides with, or is situated slightly before, the corresponding threshold error rate $p_c$.

\begin{figure*}[t]
	\centering
	\includegraphics[width=0.95\textwidth]{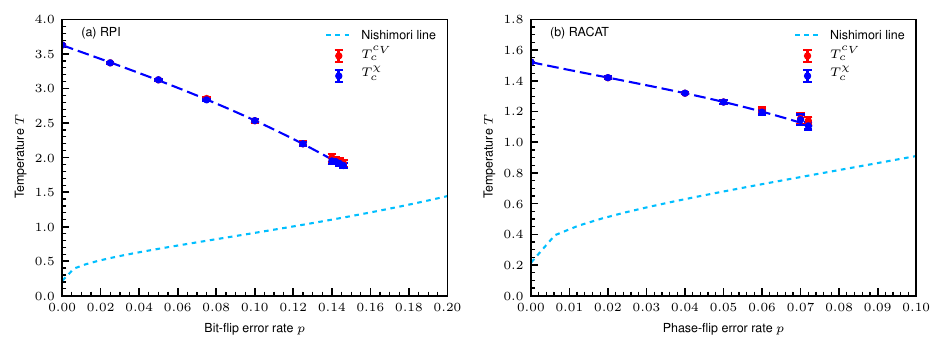}
	\caption{Phase boundaries in the region of first-order phase transitions for (a) the 3D RPI model and (b) the 3D RACAT model. $T_c$'s are estimated by fitting the non-standard scaling relation Eq.~\eqref{eq:scaling}. Results obtained from the locations of $C_v$ and $\chi$ maximum are compared.}
	\label{fig:fig_PD_chi_Cv}
\end{figure*}

\subsection{Non-standard first-order scaling}

For a typical first-order phase transition, such as that in the $2$D $q$-state Potts model with $q\geq 5$, the finite-size transition temperature $T_c(L)$ is shifted by an amount  $L^{-d}$ from $T_c = T_c(\infty)$~\cite{Challa86, Lee91}, where $d$ is the dimension of the system.
However, it has been understood recently that, in the presence of planar spin-flip symmetries such as in the disorder-free models $H_{\eta=0}^{\mathcal{A}}$ and $H_{\eta=0}^{\mathcal{B}}$, the leading order finite-size correction to $T_c$ is modified to $L^{-(d-1)}$ due to the  sub-extensive degeneracy ($\sim 2^{3L}$)~\cite{Mueller14, Johnston17}.
As the quenched disorder does not affect the subsystem symmetries of $H_{\eta}^{\mathcal{A}}$ and $H_{\eta}^{\mathcal{B}}$, and their first-order phase transitions are robust against finite $p$ (see Fig.~\ref{fig:PB_PIM} and Fig.~\ref{fig:PB_ATM}),
we expect that the modified scaling relation still holds, at least for the low $p$ regions, and we extrapolate $T_c$ by fitting
\begin{align}\label{eq:scaling}
	T_c(L) \sim T_c + bL^{-(d-1)}.
\end{align}
With increasing $p$, the first-order phase transitions gradually soften and the correlation length will eventually exceed our simulation system sizes.
Nevertheless, as shown in Fig.~\ref{fig:fig_PD_chi_Cv}, for all $p$ values showing a first-order phase transition, the estimated $T_c$ is sufficiently above the Nishimori line which defines the relevant $(p, T)$ values for an error code.
Therefore, while the precision on the transition temperatures is expected to be improved from larger system sizes (which are beyond reach however), this should only have minimal consequences in determining the error thresholds $p_c^X$ and $p_c^Z$ (see~\ref{sec:p_c}).

\begin{figure*}[t]
\centering
\includegraphics{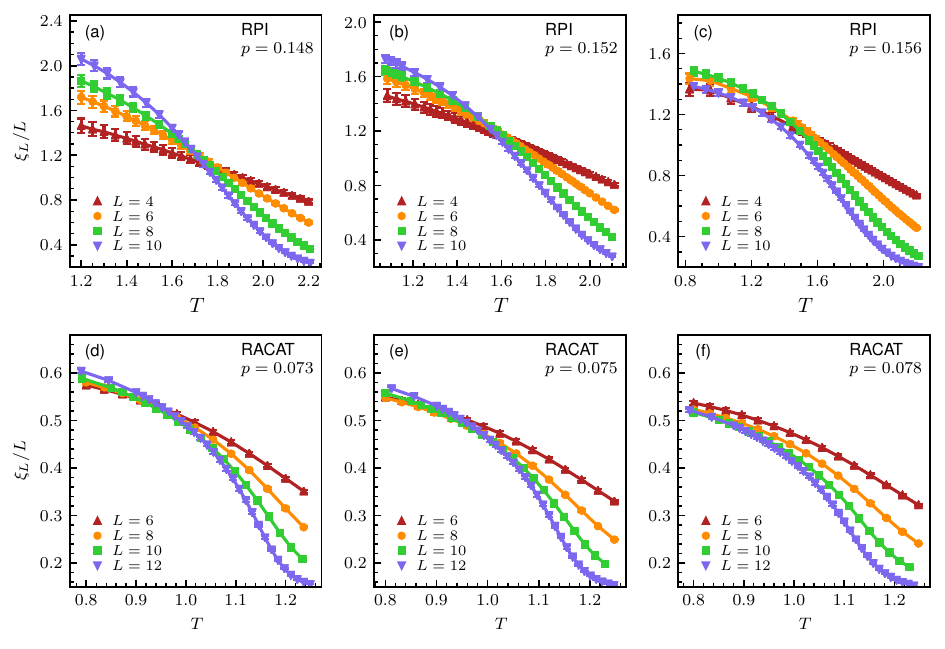}
\caption{Normalized second-moment correlation length $\xi_L/L$ for the 3D RPI [(a)-(c)] and RACAT [(d)-(f)] models in the vicinity of the estimated optimal error thresholds. A clear intersection of $\xi_L/L$ curves can be recognized up to $p_c^X \simeq 0.152(4)$ for the RPI model and up to $p_c^Z \simeq 0.075(2)$ for the RACAT model. For $p > p_c^X$ or $p_c^Z$, the intersection becomes very ambiguous or the lines do not cross, indicating no order-disorder phase transition.}
\label{fig:xi_all}
\end{figure*}

\subsection{Optimal error-threshold values}\label{sec:p_c}
In the modeling of local errors, temperature is an auxiliary variable introduced by the constraint Eq.~\eqref{eq:Nishimori2}.
Only $(p,T)$ values on the Nishimori line are relevant for quantum error correction. 
Namely, a correctable X-cube code corresponds to the part of the Nishimori line inside the ordered phase of the RPI or RACAT model.
Clearly, the high-temperature phase of both models is trivially disordered. We can thereby estimate the optimal error-threshold values $p_c^X$ and $p_c^Z$ by the largest error rates exhibiting an order-disorder phase transition.

As the phase transitions are continuous for large enough $p$ values, we use $\xi_L$ as an estimator. 
Following the scale invariance, in the vicinity of a critical point $\xi_L$ scales as
\begin{align}
	\frac{\xi_L}{L}=g\left(L^{1/\nu}\left(T-T_{c}\right)\right),
\end{align}
and $\xi_L/L$ for different sizes is expected to intersect near $T_c$, where  $g$ is a universal scaling function and $\nu$ is the critical exponent of correlation length. 
 
 We show in Figures~\ref{fig:xi_all} the curves of $\xi_L/L$ in the vicinity of the estimated optimal error thresholds.
A clear intersection can be observed at $p_c^X \simeq 0.152(4)$ for the RPI model and at $p_c^Z \simeq 0.075(2)$ for the RACAT model, indicating the existence of a second-order phase transition.
For $p > p_c^X$ or $p_c^Z$, the curves do not meet or their intersection becomes very ambiguous, and we conclude no order-disorder phase transition.

\section{Details of numerical simulations} \label{sec:simulation}
Simulating systems with quenched disorder and first-order phase transitions is generally a challenging task.
In the simulations of the 3D RPI and RACAT model, we employ parallel tempering (PT) jointly with the heat bath and over-relaxation algorithms to equilibrate the systems~\cite{Janke08}.
The distribution of temperatures is carefully chosen and tested to ensure the acceptance ratio of PT updates~\cite{Katzgraber06} for each of the disorder strengths $p$ and system sizes $L$.
Large numbers ($N_d$) of random coupling configurations are considered, with $N_d = 200$ in the low $p$ regimes and $N_d = 800, 1600$ for $p$ values near the error thresholds.
Statistical error bars are estimated by the bootstrap method~\cite{BookEfron}.
Equilibration is tested by a binning analysis.
For each simulation, the measurements are averaged over the Monte Carlo time interval $[2^{\tau-1}, 2^{\tau}-1]$, where $\tau$ labels the bins~\cite{Bombin12}.
The system is considered equilibrated when, at least, the last three bins agree within statistical uncertainty.
Simulation parameters are summarized in Table~\ref{tab:sim_PIM} and Table~\ref{tab:sim_ATM}.
The simulations ran at the CoolMUC-2 cluster and the KCS cluster at Leibniz-Rechenzentrum (LRZ) and used over three million CPU hours without taking into account the intensive tests for optimizing temperature distributions in parallel tempering.

\begin{table}[hp!]
\centering
\begin{tabular*}
{0.6\columnwidth}{@{\extracolsep{\fill} } c c c c c c c}
\hline\hline			
$p$ & $L$ & $N_{d}$ & $\tau_{\max}$ & $N_T$& $T_{\min}$ & $T_{\max}$\\
\hline
\hline
0.000 & 4, 6 & 200 & 23 & 56 & 2.50 &  6.23 \\
0.000 & 8 & 200 & 23 & 56 & 2.50 &  6.00 \\
0.000 & 10 & 200 & 22 & 56 & 3.50 &  6.00 \\
0.025 & 4, 6 & 200 & 23 & 56 & 2.00 &  5.70 \\
0.025 & 8 & 200 & 23 & 56 & 3.10 &  5.50 \\
0.025 & 10 & 200 & 22 & 56 & 3.10 &  5.50 \\
0.050 & 4, 6 & 200 & 23 & 56 & 2.00 &  5.73 \\
0.050 & 8 & 200 & 23 & 56 & 2.80 &  5.50 \\
0.050 & 10 & 200 & 22 & 56 & 2.85 &  4.50 \\
0.075 & 4, 6 & 200 & 23 & 56 & 2.00 &  5.80 \\
0.075 & 8 & 200 & 23 & 56 & 2.40 &  5.50 \\
0.075 & 10 & 200 & 22 & 56 & 2.40 &  5.50 \\
0.100 & 4, 6 & 200 & 23 & 56 & 2.00 &  5.83 \\
0.100 & 8 & 200 & 23 & 56 & 2.00 &  5.50 \\
0.100 & 10 & 200 & 22 & 56 & 2.00 &  5.00 \\
0.125 & 4, 6 & 200 & 23 & 56 & 1.70 &  5.88 \\
0.125 & 8 & 200 & 23 & 56 & 1.65 &  5.50 \\
0.125 & 10 & 200 & 22 & 56 & 1.65 &  5.36 \\
0.140 & 4, 6 & 200 & 23 & 56 & 0.30 &  5.93 \\
0.140 & 8 & 200 & 23 & 56 & 0.30 &  5.50 \\
0.140 & 10 & 200 & 22 & 56 & 1.10 &  5.00 \\
0.142 & 4, 6 & 800 & 23 & 56 & 1.30 &  5.50 \\
0.142 & 8 & 800 & 23 & 56 & 1.30 &  5.00 \\
0.142 & 10 & 800 & 22 & 56 & 1.30 &  5.00 \\
0.144 & 4, 6 & 800 & 23 & 56 & 1.15 &  5.50 \\
0.144 & 8 & 800 & 23 & 56 & 1.15 &  5.36 \\
0.144 & 10 & 800 & 22 & 56 & 1.15 &  5.36 \\
0.146 & 4, 6 & 800 & 23 & 56 & 1.10 &  5.33 \\
0.146 & 8 & 800 & 23 & 56 & 1.10 &  5.00 \\
0.146 & 10 & 800 & 22 & 56 & 1.10 &  5.00 \\
0.148 & 4, 6 & 800 & 23 & 56 & 1.10 &  5.33 \\
0.148 & 8 & 800 & 23 & 56 & 1.10 &  5.00 \\
0.148 & 10 & 800 & 22 & 56 & 1.10 &  5.00 \\
0.150 & 4, 6 & 800 & 23 & 56 & 1.00 &  5.38 \\
0.150 & 8 & 800 & 23 & 56 & 1.00 &  5.46 \\
0.150 & 10 & 800 & 22 & 56 & 1.00 &  5.46 \\
0.152 & 4, 6 & 1600 & 23 & 56 & 1.00 &  5.00 \\
0.152 & 8 & 1600 & 23 & 56 & 1.00 &  5.46 \\
0.152 & 10 & 1600 & 22 & 56 & 1.00 &  5.46 \\
0.154 & 4, 6, 8 & 1600 & 23 & 56 & 1.00 &  5.00 \\
0.154 & 10 & 1600 & 22 & 56 & 1.00 &  5.00 \\
0.156 & 4, 6, 8 & 1600 & 23 & 56 & 0.70 &  5.00\\
0.156 & 10 & 1600 & 22 & 56 & 0.70 &  5.00 \\
\hline\hline
\end{tabular*}
\caption{
Simulation parameters for the 3D RPI model. $L$ is the linear size of the system. $N_d$ denotes the number of random coupling configurations at the error rate $p$. Larger $N_d$ is considered near the $X$ error threshold $p_c^X \simeq 0.152(4)$. $2^{\tau_{\max}}$ gives the number of Monte Carlo sweeps in a simulation. $N_T$ temperatures between $T_{\min}$ and $T_{\max}$ are simulated in parallel. The same conventions are used for the 3D RACAT model in Table~\ref{tab:sim_ATM}.}
\label{tab:sim_PIM}
\end{table}

\begin{table}[hp!]
\centering
\begin{tabular*}
{0.6\columnwidth}{@{\extracolsep{\fill} } c c c c c c c}
\hline\hline			
$p$ & $L$ & $N_{d}$ & $\tau_{\max}$ & $N_T$& $T_{\min}$ & $T_{\max}$\\
\hline
0.020 & 6, 8 & 200 & 22 & 64 & 0.80 &  2.79 \\
0.020 & 10 & 200 & 22 & 64 & 1.17 &  2.29 \\
0.020 & 12 & 200 & 22 & 64 & 1.27 &  2.13 \\
0.040 & 6, 8 & 200 & 22 & 64 & 0.98 &  2.41 \\
0.040 & 10 & 200 & 22 & 64 & 1.10 &  2.14 \\
0.040 & 12 & 200 & 22 & 64 & 1.13 &  2.07 \\
0.050 & 6, 8 & 200 & 22 & 64 & 0.44 &  2.74 \\
0.050 & 10 & 200 & 22 & 64 & 0.81 &  2.37 \\
0.050 & 12 & 200 & 22 & 64 & 1.08 &  2.11 \\
0.060 & 6, 8 & 200 & 22 & 64 & 0.39 & 2.78 \\
0.060 & 10 & 200 & 22 & 64 & 0.61 &  2.36 \\
0.060 & 12 & 200 & 22 & 64 & 0.91 &  2.14 \\
0.070 & 6, 8 & 200 & 22 & 64 & 0.30 &  2.60 \\
0.070 & 10 & 200 & 22 & 64 & 0.50 &  2.45 \\
0.070 & 12 & 200 & 22 & 64 & 0.66 &  2.25 \\
0.072 & 6, 8 & 200 & 22 & 56 & 0.30 &  2.70 \\
0.072 & 10 & 200 & 22 & 56 & 0.35 &  2.50 \\
0.072 & 12 & 200 & 22 & 56 & 0.53 &  2.26 \\
0.073 & 6, 8 & 800 & 22 & 56 & 0.30 &  2.70 \\
0.073 & 10 & 800 & 22 & 56 & 0.35 &  2.50 \\
0.073 & 12 & 800 & 22 & 56 & 0.53 &  2.26 \\
0.074 & 6, 8 & 800 & 22 & 64 & 0.30 &  2.60 \\
0.074 & 10 & 800 & 22 & 64 & 0.35 &  2.50 \\
0.074 & 12 & 800 & 22 & 64 & 0.53 &  2.25 \\
0.075 & 6, 8 & 800 & 22 & 64 & 0.30 &  2.60 \\
0.075 & 10 & 800 & 22 & 64 & 0.35 &  2.50 \\
0.075 & 12 & 800 & 22 & 64 & 0.53 &  2.25 \\
0.076 & 6, 8 & 800 & 22 & 64 & 0.30 &  2.60 \\
0.076 & 10 & 800 & 22 & 64 & 0.35 &  2.50 \\
0.076 & 12 & 800 & 22 & 64 & 0.53 &  2.25 \\
0.078 & 6, 8 & 800 & 22 & 64 & 0.30 &  2.60 \\
0.078 & 10 & 800 & 22 & 64 & 0.35 &  2.50 \\
0.078 & 12 & 800 & 22 & 64 & 0.51 &  2.23 \\
\hline\hline
\end{tabular*}
\caption{Simulation parameters for the 3D RACAT model.}
\label{tab:sim_ATM}
\end{table}

\end{document}